\documentclass[aps,prb,twocolumn,notitlepage,superscriptaddress, preprintnumbers]{revtex4-2}
\usepackage{graphicx}
\usepackage{blindtext}
\usepackage{amsmath,amssymb}
\usepackage{color}
\usepackage[colorlinks=true,citecolor=blue,linkcolor=blue]{hyperref}
\usepackage{bm}
\usepackage{bbm}
\usepackage{mathtools, physics}
\usepackage{dsfont}
\usepackage{hyperref}
\usepackage{amsthm,times}
\usepackage{tcolorbox}
\usepackage[normalem]{ulem}
\theoremstyle{plain}
\newtheorem{thm}{Theorem}
\newtheorem*{thm*}{Theorem}

\usepackage{booktabs} 
\usepackage{multirow} 

\newcommand{\bR}{\bm{R}}
\newcommand{\bk}{\bm{k}}

\newcommand{\ii}{{\mathrm{i}}}

\begin{document}

\title{Frustration-free free fermions and beyond}
\author{Rintaro Masaoka}
\affiliation{Department of Applied Physics, The University of Tokyo, Tokyo 113-8656, Japan}
\author{Seishiro Ono}
\affiliation{Institute for Solid State Physics, University of Tokyo, Kashiwa 277-8581, Japan}
\affiliation{Department of Physics, Hong Kong University of Science and Technology, Clear Water Bay, Hong Kong SAR, China}
\affiliation{RIKEN Center for Interdisciplinary Theoretical and Mathematical Sciences (iTHEMS), RIKEN, Wako 351-0198, Japan}
\author{Hoi Chun Po}
\email{hcpo@ust.hk}
\affiliation{Department of Physics, Hong Kong University of Science and Technology, Clear Water Bay, Hong Kong, China}
\affiliation{Center for Theoretical Condensed Matter Physics, \\
	Hong Kong University of Science and Technology, Clear Water Bay, Hong Kong, China}
\author{Haruki Watanabe}\email{hwatanabe@g.ecc.u-tokyo.ac.jp}
\affiliation{Department of Applied Physics, The University of Tokyo, Tokyo 113-8656, Japan}
\date{\today}

\preprint{RIKEN-iTHEMS-Report-25}

\begin{abstract}
Frustration-free Hamiltonians provide pivotal models for understanding quantum many-body systems. In this paper, we establish a general framework for frustration-free fermionic systems. First, we derive a necessary and sufficient condition for a free fermion model to be frustration-free. In the case of translation-invariant, noninteracting systems, we show that any band touching between the valence and conduction bands is at least quadratic. Furthermore, by extending the Gosset-Huang inequality to fermionic systems, we demonstrate that even in interacting and non-translation-invariant cases, the finite-size gap of gapless excitations scales as $O((\log L)^2/L^2)$, provided the ground-state correlation function exhibits a power-law decay. Our results provide a foundation for studying frustration-free fermionic systems, including flat-band ferromagnetism and $\eta$-pairing states.
\end{abstract}

\maketitle

\section{Introduction}

In condensed matter physics, \textit{frustration}---the impossibility of minimizing all terms in the Hamiltonian simultaneously---is a key concept.
One of the most fundamental examples is geometrical frustration in magnets.
For example, in the antiferromagnetic Heisenberg model on a triangular lattice, no spin configuration can simultaneously minimize all antiferromagnetic exchange interactions.

In contrast, we say that a Hamiltonian is \textit{frustration-free} if the ground state of the total Hamiltonian is also a ground state of each local term in the Hamiltonian.
Although frustration-free systems may appear to form a narrow class, they contain representative models of a diverse range of quantum phases of matter, such as quantum critical points, spontaneously symmetry breaking (SSB) phases, symmetry protected topological (SPT) phases, and topologically ordered phases.
For instance, the Rokhsar-Kivelson model~\cite{PhysRevLett.61.2376}, the Majumdar--Ghosh model~\cite{10.1063/1.1664978, Majumdar:1970aa}, the Affleck--Kennedy--Lieb--Tasaki model~\cite{AKLT, TasakiBook}, and the toric code model~\cite{KITAEV20032} are frustration-free models that represent a quantum critical point, an SSB phase, an SPT phase, and a topologically ordered phase, respectively.
More generally, the parent Hamiltonians of matrix product states and projected entangled pair states are also frustration-free~\cite{parent, parent_PEPS}.

Frustration-free gapless systems exhibit features substantially different from those of frustrated systems~\cite{PhysRevLett.120.117202,PhysRevB.103.214428,PhysRevLett.131.220403,arXiv:2310.16881,arXiv:2405.00785}.
Recently, it has been conjectured that when a frustration-free gapless Hamiltonian is translationally invariant, the dispersion relation of low-energy excitations above the ground state is generally quadratic or softer~\cite{arXiv:2310.16881,arXiv:2406.06414}. This implies that the low-energy effective theory of gapless frustration-free systems cannot be Lorentz invariant.
Remarkably, such quadratic dispersions allow SSB of continuous symmetries at zero temperature even in one dimension~\cite{arXiv:2310.16881}, which is typically forbidden by the Hohenberg-Mermin-Wagner theorem~\cite{PhysRevLett.17.1133,PhysRev.158.383,RevModPhys.69.315}.
Also, when the ground state correlation function shows a power-law decay, the finite-size gap of gapless excitations scales as $O(\log L)^2/L^2)$~\cite{arXiv:2406.06415}\footnote{Under the open boundary condition, a tighter bound $O((L^{-2})$ is derived without assuming power-law decay of correlation functions~\cite{Knabe,Gossetozgunov,Anshu,Lemm_2022,lemm2024criticalfinitesizegapscaling}.}.
To date, these intriguing properties have been studied primarily in frustration-free spin systems. It is therefore natural to ask whether fermionic frustration-free systems have similar characteristics. However, despite some studies on frustration-free fermionic systems~\cite{PhysRevB.92.115137,PhysRevB.98.155119,Jevtic_2017,PhysRevResearch.3.033265,jones2023exact}, their properties remain less comprehensively explored compared to those of spin systems.

In this work, we study frustration-free fermionic systems.
First, we focus on frustration-free free fermionic systems, where Hamiltonians are expressed as quadratic forms of fermionic creation and annihilation operators.
We derive a  necessary and sufficient condition for free fermionic systems to be frustration-free. 
Such a condition has been previously investigated for 1D one-band models in class BDI~\cite{PhysRevResearch.3.033265} and 1D two-band models in class AIII~\cite{jones2023exact} from a different approach. Our results are more general in that no assumptions are made regarding  the dimensionality of the system, the number of bands, or the symmetry classes.
We find a deep connection to tight-binding models with flat bands~\cite{Sutherland,Mielke,PhysRevB.78.125104,PhysRevB.95.115309,PhysRevB.99.045107,PhysRevB.104.085144,BAB,BAB2,TasakiBook}.
Leveraging this condition and band theory, we prove that for gapless systems with translation symmetry, the excitation energy disperses as $O(|\bm{k}-\bm{k}_0|^2)$ around some $\bm{k}_0$.

We also investigate the topological properties of frustration-free free fermionic models. We establish that the valence bands in these systems always admit an (over)complete set of compactly supported Wannier-type functions, provided that the valence band energy is strictly negative. The classification of topological phases with such Wannier-type functions~\cite{PhysRevB.95.115309} implies that stable topological phases, including Chern insulators and $\mathbb{Z}_2$ topological insulators, cannot be realized in frustration-free free fermionic models. In contrast, we demonstrate via a concrete model that fragile topology~\cite{PhysRevLett.121.126402} can be realized.

We then extend our results to more general fermionic systems, including cases with interactions and without translation symmetry, by generalizing the Gosset-Huang inequality~\cite{GossetHuang} to fermionic cases.
With this extension, we show that even in interacting and/or non-translationally invariant fermionic systems, the finite-size gap of gapless excitations scales as $O((\log L)^2/L^2)$, provided the ground state correlation function exhibits a power-law decay.

The main results of this work are presented in conjunction with a companion letter~\cite{ono2025frustrationfreefreefermions}. 
While the companion letter 
includes separate discussions on a frustration-free model 
on the honeycomb lattice and the effects of interactions in that model, the present paper establishes the comprehensive general framework. Here, we provide rigorous mathematical derivations, extend the analysis to $\text{U}(1)$ symmetry-breaking systems, detail the topological properties, and prove the finite-size gap scaling in interacting systems.

The remainder of this paper is organized as follows.
In Sec.~\ref{sec:U1}, we derive a necessary and sufficient condition for frustration-freeness in free fermionic systems.
We also show that this condition implies that the low-energy excitation is quadratic or softer for frustration-free free fermionic systems with translation symmetry.
In Sec.~\ref{sec:examples}, we present several illustrative examples.
In Sec.~\ref{sec:BdG}, we discuss the case of $\text{U}(1)$-symmetry breaking.
In Sec.~\ref{sec:topologicalphases}, we examine the possible stable topology realizable within frustration-free free-fermion models.
In Sec.~\ref{sec:GH}, we discuss the Gosset--Huang inequality for general fermionic systems.

\section{U(1)-symmetric free fermionic systems}
\label{sec:U1}
In this section, we consider noninteracting fermions described by a tight-binding model with U(1) symmetry. 
\subsection{Setting}
In general, the Hamiltonian $\hat{H}=\sum_{\bm{R}\in\Lambda}\hat{H}_{\bm{R}}$ is given by a sum of local terms $\hat{H}_{\bm{R}}$, which take the quadratic form
\begin{align}
\hat{H}_{\bm{R}}&=\sum_{\bm{R}',\bm{R}''\in\Lambda}\sum_{\tau,\tau'=1}^{N_{\bm{R}}}\hat{c}_{\bm{R}'\tau}^\dagger (t_{\bm{R}})_{\bm{R}'\tau,\bm{R}''\tau'}\hat{c}_{\bm{R}''\tau'}+C_{\bm{R}}.\label{generalH}
\end{align}
Here, $C_{\bm{R}}$ is a constant that will be chosen later, $\bm{R}\in\Lambda$ is the unit cell label, and the indices $\tau,\tau'=1,2,\cdots,N_{\bm{R}}$ include all internal degrees of freedom.
The creation/annihilation operators satisfy the anti-commutation relations $\{\hat{c}_{\bm{R}\tau},\hat{c}_{\bm{R}'\tau'}^\dagger\}=\delta_{\bm{R},\bm{R}'}\delta_{\tau,\tau'}$ and $\{\hat{c}_{\bm{R}\tau},\hat{c}_{\bm{R}'\tau'}\}=0$. The Hamiltonian is local in the sense that $(t_{\bm{R}})_{\bm{R}'\tau,\bm{R}''\tau'}=0$ whenever $|\bm{R}'-\bm{R}|>r$ or $|\bm{R}''-\bm{R}|>r$. This $r$ is called the range of the hopping, which is assumed to be finite and independent of the system size. 
Hence, we can regard the hopping amplitude $(t_{\bm{R}})_{\bm{R}'\tau,\bm{R}''\tau'}$ in Eq.~\eqref{generalH} as a matrix with indices $\bm{R}'\tau$ and $\bm{R}''\tau'$ and $\hat{c}_{\bm{R}''\tau'}$ as a vector with an index $\bm{R}''\tau'$. Then the local term can be expressed as
\begin{align}
\hat{H}_{\bm{R}}&=\hat{\bm{c}}_{\bm{R}}^\dagger h_{\bm{R}}\hat{\bm{c}}_{\bm{R}}+C_{\bm{R}}.\label{generalH2}
\end{align}
We note that the frustration-free condition derived in this work has been extended to quasi-local Hamiltonians with interaction tails decaying faster than any power law~\cite{KITAEV20062,Sengoku2025}.

The general expressions in Eqs.~\eqref{generalH} and \eqref{generalH2} do not assume the translation invariance. We will restrict ourselves to translation invariant case in Sec.~\ref{TRinv}. We also discuss the $\text{U}(1)$ broken case in Sec.~\ref{sec:BdG} and interacting case  in Sec.~\ref{sec:GH}.

\subsection{Condition for frustration-freeness}
First, let us derive the condition for the Hamiltonian $\hat{H}$ to be frustration-free.  If the ground state energy of the local Hamiltonian $\hat{H}_{\bm{R}}$ is $E_{\bm{R}}$, then $\hat{H}=\sum_{\bm{R}\in\Lambda}\hat{H}_{\bm{R}}$ is frustration free if and only if there exists a state $|\Phi\rangle$ such that 
\begin{align}
\hat{H}_{\bm{R}}|\Phi\rangle=E_{\bm{R}}|\Phi\rangle.\label{definitionofFFFF}
\end{align}
for all $\bm{R}$ simultaneously.

For example, when $C_{\bm{R}}=0$ and $h_{\bm{R}}$ is positive-semidefinite (i.e., its eigenvalues are all nonnegative), then the Fock vacuum $|0\rangle$ satisfies this condition. In this sense, any tight-binding model is frustration-free as long as the chemical potential is set to sufficiently low.

To investigate the possibility of more nontrivial ground states, 
let $\mu_{\bm{R}\alpha}>0$ ($\alpha=1,2,\cdots,A_{\bm{R}}$) be positive eigenvalues of $h_{\bm{R}}$ and $\bm{\psi}_{\bm{R}\alpha}$ be the corresponding orthonormalized eigenvectors.
Also, let $-\nu_{\bm{R}\beta}<0$ ($\beta=1,2,\cdots,B_{\bm{R}}$) be negative eigenvalues and $\bm{\phi}_{\bm{R}\beta}$ be the corresponding orthonormalized eigenvectors.
There might be eigenvectors with precisely zero eigenvalues but they will not be used in the following discussions. It follows by definition that
\begin{align}
\label{eq:Hamiltonianpm}
h_{\bm{R}}&=\sum_{\alpha=1}^{A_{\bm{R}}}\mu_{\bm{R}\alpha}\bm{\psi}_{\bm{R}\alpha}\bm{\psi}_{\bm{R}\alpha}^\dagger-\sum_{\beta=1}^{B_{\bm{R}}}\nu_{\bm{R}\beta}\bm{\phi}_{\bm{R}\beta}\bm{\phi}_{\bm{R}\beta}^\dagger.
\end{align}
We define corresponding annihilation operators by 
\begin{align}
\hat{\psi}_{\bm{R}\alpha}\coloneqq\bm{\psi}_{\bm{R}\alpha}^\dagger\hat{\bm{c}}_{\bm{R}},\label{psiR}\\
\hat{\phi}_{\bm{R}\beta}\coloneqq\bm{\phi}_{\bm{R}\beta}^\dagger\hat{\bm{c}}_{\bm{R}}.\label{phiR}
\end{align}
By construction, these operators within the same unit cell $\bm{R}$ satisfy
\begin{align}
&\{\hat{\psi}_{\bm{R}\alpha},\hat{\psi}_{\bm{R}\alpha'}^\dagger\}=\delta_{\alpha,\alpha'},\quad\{\hat{\phi}_{\bm{R}\beta},\hat{\phi}_{\bm{R}\beta'}^\dagger\}=\delta_{\beta,\beta'},\label{AC1}\\
&\{\hat{\psi}_{\bm{R}\alpha},\hat{\phi}_{\bm{R}\beta}^\dagger\}=0\label{AC2}
\end{align}
for any $\alpha,\alpha',\beta,\beta'$. For different unit cells $\bm{R}'\neq\bm{R}$, $\{\hat{\psi}_{\bm{R}\alpha},\hat{\psi}_{\bm{R}'\alpha'}^\dagger\}$, $\{\hat{\phi}_{\bm{R}\beta},\hat{\phi}_{\bm{R}'\beta'}^\dagger\}$, and $\{\hat{\psi}_{\bm{R}\alpha},\hat{\phi}_{\bm{R}'\beta}^\dagger\}$ do not necessarily vanish, but the trivial relations
\begin{align}
&\{\hat{\psi}_{\bm{R}\alpha},\hat{\psi}_{\bm{R}'\alpha'}\}=\{\hat{\phi}_{\bm{R}\beta},\hat{\phi}_{\bm{R}'\beta'}\}=\{\hat{\psi}_{\bm{R}\alpha},\hat{\phi}_{\bm{R}'\beta}\}=0\label{AC3}
\end{align}
still hold.

If we set $C_{\bm{R}}=\sum_{\beta=1}^{B_{\bm{R}}}\nu_{\bm{R}\beta}$, the Hamiltonian can be rewritten as $\hat{H}_{\bm{R}}=\hat{H}_{\bm{R}}^{(+)}+\hat{H}_{\bm{R}}^{(-)}$ with
\begin{align}
\hat{H}_{\bm{R}}^{(+)}&\coloneqq\sum_{\alpha=1}^{A_{\bm{R}}}\mu_{\bm{R}\alpha}\hat{\psi}_{\bm{R}\alpha}^\dagger\hat{\psi}_{\bm{R}\alpha},\label{def of H+}\\
\hat{H}_{\bm{R}}^{(-)}&\coloneqq\sum_{\beta=1}^{B_{\bm{R}}}\nu_{\bm{R}\beta}\hat{\phi}_{\bm{R}\beta}\hat{\phi}_{\bm{R}\beta}^\dagger.\label{def of H-}
\end{align}
Note that both $\hat{\psi}_{\bm{R}\alpha}^\dagger\hat{\psi}_{\bm{R}\alpha}$ and $\hat{\phi}_{\bm{R}\beta}\hat{\phi}_{\bm{R}\beta}^\dagger=1-\hat{\phi}_{\bm{R}\beta}^\dagger\hat{\phi}_{\bm{R}\beta}$ are projectors, whose eigenvalues are either $0$ or $1$. Therefore, $\hat{H}_{\bm{R}}$ is positive semidefinite as long as $\mu_{\bm{R}\alpha}>0$ and $\nu_{\bm{R}\beta}>0$. Also,  the ground state energy of the local Hamiltonian $\hat{H}_{\bm{R}}$  is $0$, since, for each $\bm{R}$, the state $|\Phi_{\bm{R}}\rangle\coloneqq\prod_{\beta=1}^{B_{\bm{R}}}\hat{\phi}_{\bm{R}\beta}^\dagger|0\rangle$ satisfies $\hat{\psi}_{\bm{R}\alpha}|\Phi_{\bm{R}}\rangle=\hat{\phi}_{\bm{R}\beta}^\dagger|\Phi_{\bm{R}}\rangle=0$ for all $\alpha$ and $\beta$ thanks to the anti-commutation relations in Eqs.~\eqref{AC1} and \eqref{AC2}. Therefore, the Hamiltonian is frustration free if and only if there exists a state $|\Phi\rangle$ such that 
\begin{align}
\hat{\psi}_{\bm{R}\alpha}|\Phi\rangle=\hat{\phi}_{\bm{R}'\beta}^\dagger|\Phi\rangle=0.\label{definitionofFFFF2}
\end{align}
for all $\bm{R}$, $\bm{R}'$ $\alpha$, $\beta$ simultaneously. If $\hat{H}$ is frustration free, any ground state of $\hat{H}$ satisfies the relation \eqref{definitionofFFFF2}.

Now we argue that the necessary and sufficient condition for $\hat{H}$ to be frustration-free is that
\begin{align}
\{\hat{\psi}_{\bm{R}\alpha},\hat{\phi}_{\bm{R}'\beta}^\dagger\}=0\quad\text{for all $\bm{R}$, $\bm{R}'$, $\alpha$, $\beta$.}\label{conditiontobeFFFF}
\end{align}
The necessity of Eq.~\eqref{conditiontobeFFFF} can be directly seen by applying $\{\hat{\psi}_{\bm{R}\alpha},\hat{\phi}_{\bm{R}'\beta}^\dagger\}\in\mathbb{C}$ to the ground state $|\Phi\rangle$ and using Eq.~\eqref{definitionofFFFF2}.
To see the sufficiency of Eq.~\eqref{conditiontobeFFFF}, we introduce an arbitrary ordering of $\hat{\phi}_{\bm{R}\beta}^\dagger$ for all $\bm{R}\in\Lambda$ and $\beta=1,2,\cdots B_{\bm{R}}$. Let  $\hat{\phi}_i^\dagger$ ($i=1,2,\cdots,N_{\text{tot}}\coloneqq\sum_{\bm{R}\in\Lambda}B_{\bm{R}}$) be the ordered operators. We start from the Fock vacuum $|\Phi_{0}\rangle=|0\rangle$ and successively define $|\Phi_i\rangle$ by
\begin{align}
|\Phi_i\rangle\coloneqq
\begin{cases}
\frac{1}{\|\hat{\phi}_i^\dagger|\Phi_{i-1}\rangle\|}\hat{\phi}_i^\dagger|\Phi_{i-1}\rangle & \text{if $\hat{\phi}_i^\dagger|\Phi_{i-1}\rangle\neq0$}\\
|\Phi_{i-1}\rangle & \text{if $\hat{\phi}_i^\dagger|\Phi_{i-1}\rangle=0$}
\end{cases}.
\end{align}
By construction, $|\Phi\rangle=|\Phi_{N_{\text{tot}}}\rangle$ automatically satisfies Eq.~\eqref{definitionofFFFF2}.

An important consequence of Eq.~\eqref{conditiontobeFFFF} is that, when $\hat{H}=\hat{H}_{\bm{R}}^{(+)}+\hat{H}_{\bm{R}}^{(-)}$ is frustration free, $\hat{H}_{\bm{R}}^{(+)}$ and $\hat{H}_{\bm{R}'}^{(-)}$ commute:
\begin{align}
&[\hat{H}_{\bm{R}}^{(+)},\hat{H}_{\bm{R}'}^{(-)}]=\sum_{\alpha=1}^{A_{\bm{R}}}\sum_{\beta=1}^{B_{\bm{R}}}\mu_{\bm{R}\alpha}\nu_{\bm{R}\beta}\notag\\
&\times\big(\hat{\phi}_{\bm{R}'\beta}^\dagger\{\hat{\phi}_{\bm{R}'\beta},\hat{\psi}_{\bm{R}\alpha}^\dagger\}\hat{\psi}_{\bm{R}\alpha}
-
\hat{\psi}_{\bm{R}\alpha}^\dagger\{\hat{\psi}_{\bm{R}\alpha},\hat{\phi}_{\bm{R}'\beta}^\dagger\}\hat{\phi}_{\bm{R}'\beta}\big)=0.\label{RScommutation}
\end{align}
Below we will discuss the implication of Eq.~\eqref{RScommutation}.

\subsection{Translation invariant models}
\label{TRinv}
Consider a translation-invariant frustration-free free fermionic Hamiltonian $\hat{H}=\sum_{\bm{R}}\hat{H}'_{\bm{R}}$.
Let $\hat{T}_{\bm{R}'}$ denote the translation operator by a vector $\bm{R}'$. 
Note that the translation invariance of the total Hamiltonian $\hat{H}$ does not necessarily imply that the local terms $\hat{H}'_{\bm{R}}$ respect the translation symmetry, i.e., $\hat{T}_{\bm{R}'}\hat{H}'_{\bm{R}}\hat{T}_{\bm{R}'}^\dagger=\hat{H}'_{\bm{R}+\bm{R}'}$. However, we can always construct a frustration-free decomposition $\hat{H}=\sum_{\bm{R}}\hat{H}_{\bm{R}}$ that satisfies this condition by defining
\begin{align} &
\hat{H}_{\bm{R}} = \frac{1}{V}\sum_{\bm{R}'} \hat{T}_{\bm{R}'}\hat{H}'_{\bm{R}-\bm{R}'}\hat{T}_{\bm{R}'}^{\dagger},
\end{align}
where $V$ is the number of unit cells~\footnote{Note, however, that $\hat{H}_{\bm{R}}$ constructed this way may depend on $V$.}. Consequently, we may drop the subscript $\bm{R}$ from $N_{\bm{R}}$, $A_{\bm{R}}$, $B_{\bm{R}}$, $\mu_{\bm{R}\alpha}$, and $\nu_{\bm{R}\beta}$.
The tight-binding model $\hat{H}$ has $N_{\bm{R}}=N$ bands in total.
Introducing the Fourier transformation by $\hat{c}_{\bm{R}\tau}^\dagger=\sum_{\bm{k}}\frac{1}{\sqrt{V}}e^{-i\bm{k}\cdot\bm{R}}\hat{c}_{\bm{k}\tau}^\dagger$ ($\tau=1,2,\cdots,N$), we find
\begin{align}
&\sum_{\bm{R}\in\Lambda}\hat{H}_{\bm{R}}^{(+)}=\sum_{\bm{k}}\hat{H}_{\bm{k}}^{(+)}=\sum_{\bm{k}}\hat{\bm{c}}_{\bm{k}}^\dagger H_{\bm{k}}^{(+)}\hat{\bm{c}}_{\bm{k}},\label{def of H+ matrix}\\
&\sum_{\bm{R}\in\Lambda}\hat{H}_{\bm{R}}^{(-)}=\sum_{\bm{k}}\hat{H}_{\bm{k}}^{(-)}=\sum_{\bm{k}}\Big(\hat{\bm{c}}_{\bm{k}}^\dagger H_{\bm{k}}^{(-)}\hat{\bm{c}}_{\bm{k}}+\sum_{\beta=1}^{B}\nu_{\beta}\Big).\label{def of H- matrix}
\end{align}
Here, $\hat{\bm{c}}_{\bm{k}}^\dagger=(\hat{c}_{\bm{k}1}^\dagger,\cdots,\hat{c}_{\bm{k}N}^\dagger)$ is the $N$-component row vector and $H_{\bm{k}}^{(+)}$ and $H_{\bm{k}}^{(-)}$ are $N$-dimensional Hermitian matrices.
The commutation relation in Eq.~\eqref{RScommutation} implies $[H_{\bm{k}}^{(+)},H_{\bm{k}}^{(-)}]=0$.
Therefore, the band structure of the $N$-dimensional matrix $H_{\bm{k}}\coloneqq H_{\bm{k}}^{(+)}+H_{\bm{k}}^{(-)}$ can be computed by diagonalizing $H_{\bm{k}}^{(+)}$ and $H_{\bm{k}}^{(-)}$ separately.
This implies that the band dispersions of $H_{\bm{k}}^{(+)}$ and $H_{\bm{k}}^{(-)}$, by themselves, are smooth along any one-dimensional curve $\bm{k}(t)$ in the Brillouin zone (see Appendix ~\ref{app:analytic}).

Among the eigenvalues of $H_{\bm{k}}^{(+)}$, suppose $\omega_{\bm{k}n}^{(+)}\geq0$ ($n=1,2,\cdots,N^{(+)}$) are not identically zero on the entire Brillouin zone. 
Namely, $\omega_{\bm{k}n}^{(+)}>0$ at some $\bm{k}$. 
Similarly, the eigenvalues $\omega_{\bm{k}m}^{(-)}\leq0$ ($m=1,2,\cdots,N^{(-)}$) of $H_{\bm{k}}^{(-)}$ are not identically zero.  
As a result, other $N-N^{(+)}-N^{(-)} \geq 0$ eigenvalues are completely $0$.
In general, the numbers of bands are bounded as
\begin{align}
N^{(+)}\leq A,\quad N^{(-)}\leq B,\quad N^{(+)}+N^{(-)}\leq N.
\end{align}
A ground state of $\hat{H}$ can be obtained by fully occupying all negative energy states $\omega_{\bm{k}m}^{(-)}<0$ and leaving all positive energy states $\omega_{\bm{k}n}^{(+)}>0$ completely unoccupied. Hence,
\begin{align}
\sum_{\beta=1}^{B}\nu_{\beta}=-\frac{1}{|\Lambda|}\sum_{\bm{k}}\sum_{m=1}^{N^{(-)}}\omega_{\bm{k}m}^{(-)}.
\end{align}
Zero-energy states, if exist, can be either occupied or unoccupied. Hence, if there are $N_0$ zero-energy states, the ground state degeneracy is $2^{N_0}$-fold. If a flat band exists, the ground state degeneracy is exponentially large.

The system is gapless if and only if at least one of the bands touches the zero-energy level. For example, suppose $\omega_{\bm{k}m}^{(+)}=0$ at $\bm{k}=\bm{k}_0^{(+)}$. Then the analyticity of $\omega_{\bm{k}m}^{(+)}$ combined with $\omega_{\bm{k}m}^{(+)}\geq 0$ implies that the form of band dispersion around $\bm{k}=\bm{k}_0^{(+)}$ is quadratic or softer, i.e., $\omega_{\bm{k}m}^{(+)}=O(|\bm{k}-\bm{k}_0^{(+)})|^2)$. The same argument applies to $\omega_{\bm{k}m}^{(-)}\leq 0$. Namely, when $\omega_{\bm{k}m}^{(-)}=0$ at $\bm{k}=\bm{k}_0^{(-)}$, then  $-\omega_{\bm{k}m}^{(-)}=O(|\bm{k}-\bm{k}_0^{(-)})|^2)$.

\subsection{Flat bands}
If $N^{(0)}\coloneqq N-(N^{(+)}+N^{(-)})>0$, there exist $N^{(0)}$ zero-energy flat bands in the band structure.  Let $\bm{w}_{\bm{k}\ell}$ ($\ell=1,2,\cdots,N^{(0)}$) be the eigenvector satisfying
\begin{align}
H_{\bm{k}}\bm{w}_{\bm{k}\ell}=\bm{0}.\label{FBk}
\end{align}
Since the Hamiltonian contains only finite-range hoppings, $H_{\bm{k}}$ is a Laurent polynomial (i.e., including both positive and negative powers) of $e^{i\bm{k}\cdot\bm{a}_i}$ ($i=1,\cdots,d$), where $\bm{a}_1,\cdots ,\bm{a}_d$ are the primitive lattice vectors~\cite{PhysRevB.95.115309,PhysRevB.99.045107,PhysRevB.104.085144}. Thus the eigenvector $\bm{w}_{\bm{k}\ell}$ in Eq.~\eqref{FBk} can also be chosen as a Laurent polynomial, as long as the normalization is not required.  As a consequence, the corresponding Wannier-like state in real space
\begin{align}
\hat{w}_{\bm{R}\ell}^\dagger&\coloneqq\sum_{\bm{k}}\frac{1}{\sqrt{V}}e^{-i\bm{k}\cdot\bm{R}}\sum_{\tau=1}^N(\bm{w}_{\bm{k}\ell})_\tau\hat{\bm{c}}_{\bm{k}\tau}^\dagger\notag\\
&=\sum_{\bm{d}\in\Lambda}\sum_{\tau=1}^N\Big(\sum_{\bm{k}}\frac{1}{V}e^{i\bm{k}\cdot\bm{d}}(\bm{w}_{\bm{k}\ell})_\tau\Big)
\hat{c}_{\bm{R}+\bm{d}\,\tau}^\dagger
\end{align}
is compactly supported around the unit cell $\bm{R}$~\cite{PhysRevB.95.115309,PhysRevB.99.045107,PhysRevB.104.085144}. 
Note that, in general, these states are not orthonormal.

Since $\hat{w}_{\bm{R}\ell}^\dagger$ creates a superposition of zero energy states, $[\hat{H},\hat{w}_{\bm{R}\ell}^\dagger]=0$, which implies
\begin{align}
0&=\langle\Phi|\hat{w}_{\bm{R}\ell}\hat{H}_{\bm{R}'}^{(+)}\hat{w}_{\bm{R}\ell}^\dagger|\Phi\rangle=\sum_{\alpha=1}^A\mu_\alpha\big|\{\hat{\psi}_{\bm{R}'\alpha},\hat{w}_{\bm{R}\ell}^\dagger\}
\big|^2,\\
0&=\langle\Phi|\hat{w}_{\bm{R}\ell}^\dagger\hat{H}_{\bm{R}'}^{(-)}\hat{w}_{\bm{R}\ell}|\Phi\rangle=\sum_{\beta=1}^B\nu_\beta\big|\{\hat{\phi}_{\bm{R}'\beta}^\dagger,\hat{w}_{\bm{R}\ell}\}
\big|^2
\end{align}
for state $|\Phi\rangle$ that satisfies Eq.~\eqref{definitionofFFFF2}.
Hence, 
\begin{align}
\{\hat{\psi}_{\bm{R}'\alpha},\hat{w}_{\bm{R}\ell}^\dagger\}=\{\hat{\phi}_{\bm{R}'\beta},\hat{w}_{\bm{R}\ell}^\dagger\}=0.\label{psiw}
\end{align}

The anti-commutation relations in Eq.~\eqref{psiw} allows one to add $\hat{\tilde{H}}_{\bm{R}}^{(+)}=\sum_{\ell=1}^{N^{(0)}}\tilde{\mu}_\ell \hat{w}_{\bm{R}\ell}^\dagger\hat{w}_{\bm{R}\ell}$ or $\hat{\tilde{H}}_{\bm{R}}^{(-)}=\sum_{\ell=1}^{N^{(0)}}\tilde{\nu}_\ell \hat{w}_{\bm{R}\ell}\hat{w}_{\bm{R}\ell}^\dagger$ to the Hamiltonian without breaking the frustration-free nature of the system. Conversely, one may take the limit $\nu_{\beta}\to +0$ to convert the negative energy bands to zero energy flat bands. Therefore, frustration-free free fermionic systems with nontrivial ground states (i.e., when the Fock vacuum  is not one of the ground states) are always connected to a tight-binding model with flat bands in some limit.

Frustration-free models with flat bands have been studied in the context of flat-band ferromagnets~\cite{Mielke2,Mielke3,Tasaki,TasakiBook}. For concreteness, let us introduce spin degrees of freedom $\sigma,\sigma'={\uparrow},{\downarrow}$ in addition to $\tau,\tau'$ in Eq.~\eqref{generalH} and consider a spin-independent hopping together with the onsite interaction:
\begin{align}
\hat{H}_{\bm{R}}&=\sum_{\sigma=\uparrow,\downarrow}\hat{c}_{\bm{R}\tau\sigma}^\dagger (t_{\bm{R}})_{\bm{R}'\tau,\bm{R}''\tau'}\hat{c}_{\bm{R}'\tau'\sigma}\notag\\
&\quad+U\sum_{\tau,\tau'=1}^{N_{\bm{R}}}\hat{c}_{\bm{R}\tau \uparrow}^\dagger \hat{c}_{\bm{R}\tau \uparrow}\hat{c}_{\bm{R}\tau \downarrow}^\dagger \hat{c}_{\bm{R}\tau \downarrow}.
\end{align}
Suppose that $\hat{H}_{\bm{R}}$ is positive semidefinite and $\sum_{\bm{R}\in\Lambda}\hat{H}_{\bm{R}}$ at $U=0$ has a flat band.
Let $|\Phi_0\rangle$ be the fully polarized state in which all zero-energy single-particle states are occupied by spin-up electrons. Then, for any $U>0$, $|\Phi_0\rangle$ and its SU(2)-rotated version $e^{\alpha \hat{S}^-}|\Phi_0\rangle$ ($\alpha\in\mathbb{C}$) are common zero-energy ground state among all $\hat{H}_{\bm{R}}$'s. Furthermore, these states are the only  ground states of $\sum_{\bm{R}\in\Lambda}\hat{H}_{\bm{R}}$ if and only if the correlation function $\langle\Phi_0|\hat{c}_{\bm{R}\tau\uparrow}^\dagger\hat{c}_{\bm{R}'\tau'\uparrow}|\Phi_0\rangle$ satisfies a connectivity condition~\cite{TasakiBook}. 

Even when a negative dispersion is introduced to the flat band, the model may remain frustration-free and the ferromagnetic state might be the common ground state of all $\hat{H}_{\bm{R}}$'s for sufficiently strong $U$~\cite{Tasaki2}.

\subsection{Frustration-freeness in momentum space}
\label{sec:F4_momentum_space}
The Fourier transforms of Eqs.~\eqref{def of H+} and \eqref{def of H-} yield
\begin{align}
H^{(+)}_{\bm{k}}
&= \sum_{\alpha}
[\sqrt{\mu_{\alpha}}\bm{\psi}_{\alpha}(e^{\ii \bm{k}\cdot\bm{a}_1},\ldots,e^{\ii \bm{k}\cdot\bm{a}_d})]
 \nonumber\\ & \quad\quad
\times[\sqrt{\mu_{\alpha}}\bm{\psi}_{\alpha}(e^{\ii \bm{k}\cdot\bm{a}_1},\ldots,e^{\ii \bm{k}\cdot\bm{a}_d})]^\dagger,
\label{sum of square H+}
\\
-H^{(-)}_{\bm{k}}
&= \sum_{\beta}
[\sqrt{\nu_{\beta}}\bm{\phi}_{\beta}(e^{\ii \bm{k}\cdot\bm{a}_1},\ldots,e^{\ii \bm{k}\cdot\bm{a}_d})]
 \nonumber\\ & \quad\quad
\times[\sqrt{\nu_{\beta}}\bm{\phi}_{\beta}(e^{\ii \bm{k}\cdot\bm{a}_1},\ldots,e^{\ii \bm{k}\cdot\bm{a}_d})]^\dagger,
\label{sum of square H-}
\end{align}
where the matrices $H^{(+)}_{\bm{k}}$ and $H^{(-)}_{\bm{k}}$ are defined in Eqs.~\eqref{def of H+ matrix} and \eqref{def of H- matrix} and $\bm{\psi}_{\alpha}$ and $\bm{\phi}_{\beta}$ are finite-degree vector polynomials in $e^{\ii \bm{k}\cdot\bm{a}_j}$ ($j=1,2,\ldots,d$) defined by the Fourier transforms of Eqs.~\eqref{psiR} and \eqref{phiR}.
Note that the above equations are identities of Laurent polynomials, not just an equality on the discrete momentum points in the Brillouin zone, if the system size is sufficiently large~\footnote{If the side-lengths of the system are more than twice of the maximum width of local orbials $\hat{\psi}_{\bm{R}\alpha}$ and $\hat{\phi}_{\bm{R}\beta}$, then the momentum points are dense enough to determine polynomials in Eqs.~\eqref{sum of square H+} and \eqref{sum of square H-} uniquely.}.
Thus, frustration-freeness requires that positive semidefinite Laurent polynomials $H^{(+)}_{\bm{k}}$ and $-H^{(-)}_{\bm{k}}$ can be expressed as sums of squares of finite-degree polynomials.
Conversely, if there exist finite-degree vector polynomials $\bm{\psi}_{\alpha}$ and $\bm{\phi}_{\beta}$ such that Eqs.~\eqref{sum of square H+} and \eqref{sum of square H-} hold, we can define local orbitals $\hat{\psi}_{\bm{R}\alpha}$ and $\hat{\phi}_{\bm{R}\beta}$ by the inverse Fourier transforms of Eqs.~\eqref{psiR} and \eqref{phiR}.
Therefore, frustration-freeness in translation-invariant free fermionic systems is equivalent to the existence of sum-of-squares decompositions of Laurent polynomials $\pm H^{(\pm)}_{\bm{k}}$.

While finding such decompositions is not a simple task in general, it is necessary that $H^{(\pm)}_{\bm{k}}$ are at least Laurent polynomials for the system to be frustration-free.
This necessary condition can be used to rule out frustration-freeness in most cases.

A natural question is whether positive semidefinite Laurent polynomials $\pm H^{(\pm)}_{\bm{k}}$ always imply the existence of sum-of-squares decompositions as in Eqs.~\eqref{sum of square H+} and \eqref{sum of square H-} and hence frustration-freeness.
Actually, this question has been extensively studied in mathematics. In one or two dimensions, any positive semidefinite Laurent polynomial can be expressed as a sum of squares of polynomials~\cite{ephremidzeSimpleProofMatrixValued2009, dritschelFactoringNonnegativeOperator2025}, while in three or higher dimensions $d\geq 3$, there are counterexamples~\cite{scheidererSumsSquaresRegular1999}.
Therefore, in $d=1,2$, the necessary condition above is also sufficient for frustration-freeness, while in $d\geq 3$, there are cases where the necessary condition holds but the system is not frustration-free.

\section{Examples}
\label{sec:examples}
In this section, we discuss examples of frustration-free free fermionic systems and confirm that the condition~\eqref{conditiontobeFFFF} works for these examples.
\subsection{Ladder}
\begin{figure}[t]
\begin{center}
\includegraphics[width=\columnwidth]{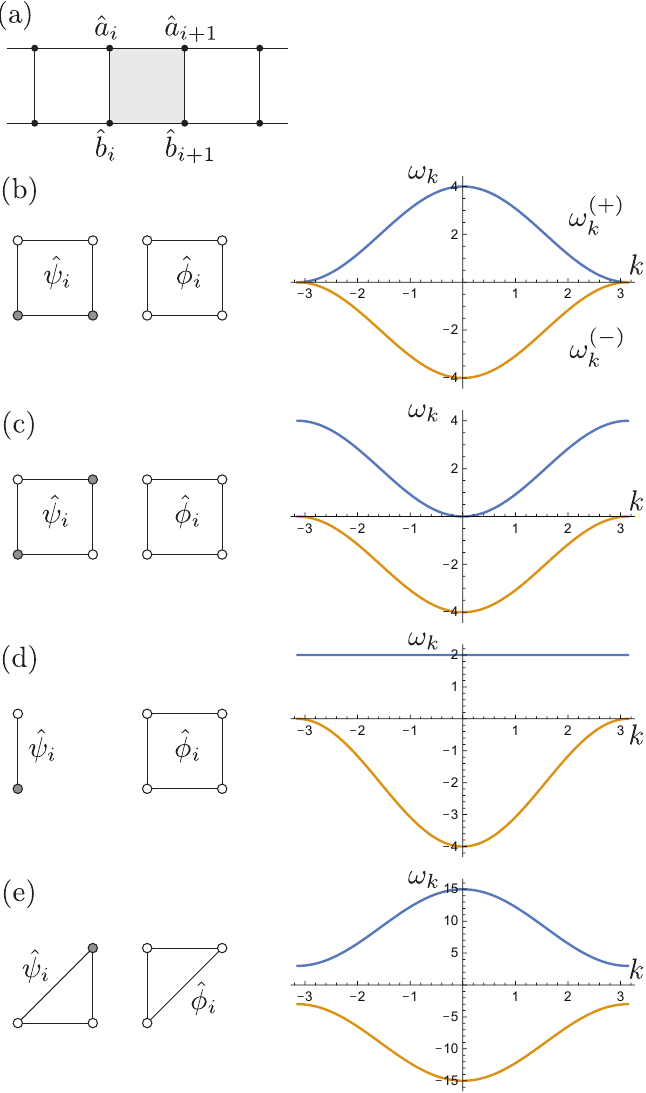}
\end{center}
\caption{
(a) Lattice structure of the ladder.
(b--e) Local orbitals and the band structure for the TB models in 
Eq.~\eqref{Ladder1} [(b)],
Eq.~\eqref{Ladder2} [(c)], 
Eq.~\eqref{Ladder3} [(d)], and 
Eq.~\eqref{Ladder4} [(f)]. White (gray) dots represent positive (negative) coefficients.
\label{figLadder}
}
\end{figure}

As simplest examples, let us discuss tight-binding models on a ladder. We write
\begin{align}
\hat{H}=\sum_{i=1}^L\hat{H}_i=\sum_{i=1}^L\bm{c}_i^\dagger h\bm{c}_i+C_i.
\end{align}
with $\bm{c}_i=(\hat{a}_i,\hat{b}_i,\hat{a}_{i+1},\hat{b}_{i+1})^\top$. See Fig.~\ref{figLadder}(a) for the illustration. The chemical potential is always set to be 0.

First, let us consider the case
\begin{align}
h=-t
\begin{pmatrix}
0&1&0&1\\
1&0&1&0\\
0&1&0&1\\
1&0&1&0\\
\end{pmatrix}
\label{Ladder1}.
\end{align}
By diagonalizing this matrix and setting  $C_i=2t$, we find $\hat{H}_i=2t\hat{\psi}_i^\dagger\hat{\psi}_i+ 2t\hat{\phi}_i\hat{\phi}_i^\dagger$ with
\begin{align}
\hat{\psi}_i&=\frac{\hat{a}_i-\hat{b}_i+\hat{a}_{i+1}-\hat{b}_{i+1}}{2},\\
\hat{\phi}_i&=\frac{\hat{a}_i+\hat{b}_i+\hat{a}_{i+1}+\hat{b}_{i+1}}{2}.
\end{align}
These orbitals, illustrated in Fig.~\ref{figLadder}(b), satisfy $\{\hat{\psi}_i,\hat{\phi}_{j}^\dagger\}=0$ for any $i,j$ and the model is frustration-free. The two bands $\omega_{k}^{(+)}=2t(1+\cos k)$ and $\omega_{k}^{(-)}=-2t(1+\cos k)$ touch at $k=\pi$ quadratically [Fig.~\ref{figLadder}(b)].

The gapless frustration-free model does not always have a quadratic band touching. For example, for the matrix
\begin{align}
h=-t
\begin{pmatrix}
0&1&1&0\\
1&0&0&1\\
1&0&0&1\\
0&1&1&0\\
\end{pmatrix}
\label{Ladder2},
\end{align}
the corresponding Hamiltonian can be rewritten as $\hat{H}_i=2t\hat{\psi}_i^\dagger\hat{\psi}_i+2t\hat{\phi}_i\hat{\phi}_i^\dagger$, where
\begin{align}
\hat{\psi}_i&=\frac{\hat{a}_i-\hat{b}_i-\hat{a}_{i+1}+\hat{b}_{i+1}}{2},\\
\hat{\phi}_i&=\frac{\hat{a}_i+\hat{b}_i+\hat{a}_{i+1}+\hat{b}_{i+1}}{2}.
\end{align}
The two bands $\omega_{k}^{(+)}=2t(1-\cos k)$ and $\omega_{k}^{(-)}=-2t(1+\cos k)$ do not touch but the indirect gap between $k=0$ and $k=\pi$ vanishes. Thus the charge-neutral sector remains gapless with quadratic dispersion [Fig.~\ref{figLadder}(c)].

Furthermore, for
\begin{align}
h=-t
\begin{pmatrix}
0&2&1&1\\
2&0&1&1\\
1&1&1&1\\
1&1&1&1\\
\end{pmatrix}
\label{Ladder3},
\end{align}
we obtain $\hat{H}_i=2t\hat{\psi}_i^\dagger\hat{\psi}_i+4t\hat{\phi}_i\hat{\phi}_i^\dagger$ with
\begin{align}
\hat{\psi}_i&=\frac{\hat{a}_i-\hat{b}_i}{\sqrt{2}},\\
\hat{\phi}_i&=\frac{\hat{a}_i+\hat{b}_i+\hat{a}_{i+1}+\hat{b}_{i+1}}{2}.
\end{align}
In this case, the upper band $\omega_{k}^{(+)}=2t>0$ is nondispersive and does not vanish at any $k$, while the lower band is dispersive and vanishes at $k=\pi$ [Fig.~\ref{figLadder}(d)]. In the ground state in which all states in the lower band are occupied, all charge neutral excitations are gapped but charged excitations with one less electron are gapless and have a quadratic dispersion. There are also gapless charge-neutral excitations if one starts with the other ground state in which the $k=\pi$ state is unoccupied.


Finally, as an example of gapped frustration-free tight-binding models, we consider the case
\begin{align}
h
= -t
\begin{pmatrix}
1&1&1&0\\
1&0&2&-1\\
1&2&0&1\\
0&-1&1&-1
\end{pmatrix}
\label{Ladder4}
\end{align}
We have $\hat{H}_i=(3t+\mu)\hat{\psi}_i^\dagger\hat{\psi}_i+(3t+\mu)\hat{\phi}_i\hat{\phi}_i^\dagger$, where
\begin{align}
\hat{\psi}_i&=\frac{\hat{b}_i-\hat{a}_{i+1}+\hat{b}_{i+1}}{\sqrt{3}},\\
\hat{\phi}_i&=\frac{\hat{a}_i+\hat{b}_i+\hat{a}_{i+1}}{\sqrt{3}}.
\end{align}
This time, a band gap opens between the two bands $\omega_{k}^{(+)}=3t(3+2\cos k)$ and $\omega_{k}^{(-)}=-3t(3+2\cos k)$ and both of these bands never touch the zero energy level  [Fig.~\ref{figLadder}(e)].

\subsection{Checkerboard lattice}
\label{sec:Checkerboard lattice}
\begin{figure}[t]
\begin{center}
\includegraphics[width=\columnwidth]{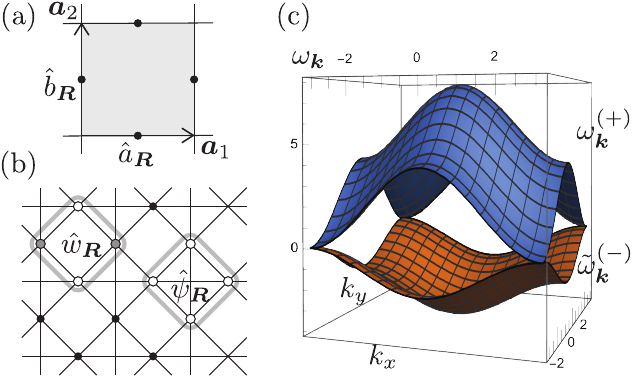}
\end{center}
\caption{
(a) The checkerboard  lattice.
(b) Local orbitals $\hat{\psi}_{\bm{R}}$ in Eq.~\eqref{CBpsi} and $\hat{w}_{\bm{R}}$ in Eq.~\eqref{CBw}.
(c) The dispersion relation with the additional term in Eq.~\eqref{CBaddition} for $t=1$ and $\tilde{\nu}=2$.
\label{figCheckerboard}
}
\end{figure}

The ladder model can be readily generalized to two dimensions. 
We consider the checkerboard lattice illustrated in Fig.~\ref{figCheckerboard} (a). 
If the nearest-neighbor bopping is $t>0$ and the onsite potential is $\mu$, the Hamiltonian is given by $\hat{H}=\sum_{\bm{R}\in\Lambda}\hat{H}_{\bm{R}}=\sum_{\bm{R}\in\Lambda}\hat{\bm{c}}_{\bm{R}}^\dagger h \hat{\bm{c}}_{\bm{R}}$ where $\hat{\bm{c}}_{\bm{R}}^\dagger=(\hat{a}_{\bm{R}}^\dagger,\hat{b}_{\bm{R}}^\dagger,\hat{a}_{\bm{R}-\bm{a}_1}^\dagger,\hat{b}_{\bm{R}-\bm{a}_2}^\dagger)$ and 
\begin{align}
h=t
\begin{pmatrix}
0&1&1&1\\
1&0&1&1\\
1&1&0&1\\
1&1&1&0\\
\end{pmatrix}+\frac{\mu}{2}
\begin{pmatrix}
1&0&0&0\\
0&1&0&0\\
0&0&1&0\\
0&0&0&1\\
\end{pmatrix}.
\end{align}
When $\mu=2t$, we find $\hat{H}_{\bm{R}}=\hat{H}_{\bm{R}}^{(+)}=4t\hat{\psi}_{\bm{R}}^\dagger\hat{\psi}_{\bm{R}}$ and 
\begin{align}
\hat{\psi}_{\bm{R}}&\coloneqq\frac{1}{2}(\hat{a}_{\bm{R}}+\hat{b}_{\bm{R}}+\hat{a}_{\bm{R}-\bm{a}_1}+\hat{b}_{\bm{R}-\bm{a}_2}).\label{CBpsi}
\end{align}
See Fig.~\ref{figCheckerboard} (b) for illustration. The Hamiltonian in the momentum space reads $\hat{H}=\sum_{\bm{k}}\hat{\bm{c}}_{\bm{k}}^\dagger H_{\bm{k}}\hat{\bm{c}}_{\bm{k}}$ with
$\hat{\bm{c}}_{\bm{k}}^\dagger=(\hat{a}_{\bm{k}}^\dagger,\hat{b}_{\bm{k}}^\dagger)$ and
\begin{align}
H_{\bm{k}}=t\begin{pmatrix}
2+2\cos k_1&(1+e^{ik_1})(1+e^{-ik_2})\\
(1+e^{-ik_1})(1+e^{ik_2})&2+2\cos k_2\\
\end{pmatrix},
\end{align}
which gives one dispersive band $\omega_{\bm{k}}^{(+)}=2t(2+\cos k_1+\cos k_2)$ and $\omega_{\bm{k}}^{(0)}=0$.
The flat band touches quadratically to the dispersive band  $\omega_{\bm{k}}^{(+)}$ at the corner of the Brillouin zone $k_1=k_2=\pi$.

The unnormalized Bloch state for the flat band is
\begin{align}
\bm{w}_{\bm{k}}=\frac{1}{2}
\begin{pmatrix}
1+e^{-ik_2}\\
-1-e^{-ik_1}\\
\end{pmatrix},\label{CBBloch}
\end{align}
which corresponds to the state $\hat{w}_{\bm{R}}$ illustrated in Fig.~\ref{figCheckerboard} (b):
\begin{align}
\hat{w}_{\bm{R}}&=\frac{1}{2}(\hat{a}_{\bm{R}}-\hat{b}_{\bm{R}}+\hat{a}_{\bm{R}+\bm{a}_2}-\hat{b}_{\bm{R}+\bm{a}_1}).\label{CBw}
\end{align}
One may also add a term
\begin{align}
\hat{\tilde{H}}^{(-)}&=\sum_{\bm{R}\in\Lambda}\tilde{\nu}\hat{w}_{\bm{R}}\hat{w}_{\bm{R}}^\dagger,\label{CBaddition}
\end{align}
which produces a negative dispersion $\tilde{\omega}_{\bm{k}}^{(-)}=-(\tilde{\nu}/2)(2+\cos k_1+\cos k_2)$ to the flat band [Fig.~\ref{figCheckerboard} (c)].

\subsection{Kagome lattice}
\begin{figure}[t]
\begin{center}
\includegraphics[width=\columnwidth]{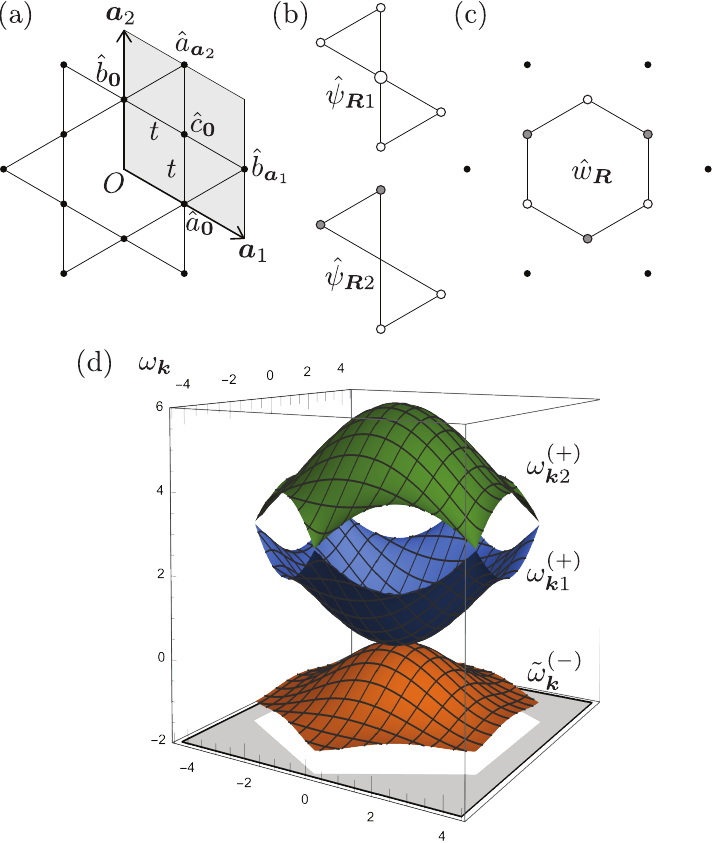}
\end{center}
\caption{
(a) The Kagome lattice.
(b) The local orbitals $\hat{\psi}_{\bm{R}1}$ and $\hat{\psi}_{\bm{R}2}$ in Eqs.~\eqref{Kagomepsi1} and \eqref{Kagomepsi2}.
(c) The hexagonal state $\hat{w}_{\bm{R}}$ in Eq.~\eqref{Kagomew}.
(d) The band dispersion for $t=\tilde{\nu}=1$.
\label{figKagome}
}
\end{figure}

As a more nontrivial model, let us discuss the tight-binding model of spinless fermions on the Kagome lattice. 
We set $\bm{a}_1=(\tfrac{\sqrt{3}}{2},-\tfrac{1}{2})$ and $\bm{a}_2=(0,1)$ (Fig.~\ref{figKagome} (a)).

If the nearest-neighbor bopping is $t>0$ and the onsite potential is $\mu$, the Hamiltonian can be written as $\hat{H}=\sum_{\bm{R}=\Lambda}\hat{\bm{c}}_{\bm{R}}^\dagger h\hat{\bm{c}}_{\bm{R}}$ where $\hat{\bm{c}}_{\bm{R}}^\dagger=(\hat{a}_{\bm{R}}^\dagger,\hat{b}_{\bm{R}+\bm{a}_1}^\dagger,\hat{c}_{\bm{R}}^\dagger,\hat{a}_{\bm{R}+\bm{a}_2}^\dagger,\hat{b}_{\bm{R}}^\dagger)$ and
\begin{align}
h=t
\begin{pmatrix}
0&1&1&0&0\\
1&0&1&0&0\\
1&1&0&1&1\\
0&0&1&0&1\\
0&0&1&1&0\\
\end{pmatrix}
+\frac{\mu}{2}\begin{pmatrix}
1&0&0&0&0\\
0&1&0&0&0\\
0&0&2&0&0\\
0&0&0&1&0\\
0&0&0&0&1\\
\end{pmatrix}.
\end{align}
When $\mu=2t$, we find $\hat{H}_{\bm{R}}=\hat{H}_{\bm{R}}^{(+)}=\sum_{\alpha=1,2}\mu_{\alpha}\hat{\psi}_{\bm{R}\alpha}^\dagger\hat{\psi}_{\bm{R}\alpha}$ with $\mu_{1}=4t$, $\mu_{2}=2t$ and
\begin{align}
\hat{\psi}_{\bm{R}1}&=\frac{1}{2\sqrt{2}}(\hat{a}_{\bm{R}}+\hat{b}_{\bm{R}+\bm{a}_1}+2\hat{c}_{\bm{R}}+\hat{a}_{\bm{R}+\bm{a}_2}+\hat{b}_{\bm{R}}),\label{Kagomepsi1}\\
\hat{\psi}_{\bm{R}2}&=\frac{1}{2}(\hat{a}_{\bm{R}}+\hat{b}_{\bm{R}+\bm{a}_1}-\hat{a}_{\bm{R}+\bm{a}_2}-\hat{b}_{\bm{R}}).\label{Kagomepsi2}
\end{align}

The Hamiltonian in the momentum space reads $\hat{H}=\sum_{\bm{k}}\hat{\bm{c}}_{\bm{k}}^\dagger H_{\bm{k}}\hat{\bm{c}}_{\bm{k}}$ with
$\hat{\bm{c}}_{\bm{k}}^\dagger=(\hat{a}_{\bm{k}}^\dagger,\hat{b}_{\bm{k}}^\dagger,\hat{c}_{\bm{k}}^\dagger)$ and
\begin{align}
H_{\bm{k}}=H_{\bm{k}}^{(+)}=\begin{pmatrix}
2&e^{ik_1}+e^{-ik_2}&1+e^{-ik_2}\\
e^{-ik_1}+e^{ik_2}&2&1+e^{-ik_1}\\
1+e^{ik_2}&1+e^{ik_1}&2
\end{pmatrix}.
\end{align}
The band dispersions are
\begin{align}
&\omega_{\bm{k}1}^{(+)}= 3t-t\sqrt{9-2e_{\bm{k}}},\\
&\omega_{\bm{k}2}^{(+)}= 3t+t\sqrt{9-2e_{\bm{k}}},\\
&\omega_{\bm{k}}^{(0)}= 0.
\end{align}
where $e_{\bm{k}}\coloneqq 3-\cos k_1-\cos k_2-\cos(k_1+k_2)\geq0$.
The unnormalized Bloch state for the flat band is
\begin{align}
\bm{w}_{\bm{k}}=\frac{1}{\sqrt{6}}
\begin{pmatrix}
1-e^{ik_1}\\
1-e^{ik_2}\\
-1+e^{i(k_1+k_2)}\\
\end{pmatrix},
\end{align}
which corresponds to the hexagonal state illustrated in Fig.~\ref{figKagome} (c).
\begin{align}
\hat{w}_{\bm{R}}&=\frac{1}{\sqrt{6}}(\hat{a}_{\bm{R}}-\hat{c}_{\bm{R}}+\hat{b}_{\bm{R}}-\hat{a}_{\bm{R}-\bm{a}_1}+\hat{c}_{\bm{R}-\bm{a}_1-\bm{a}_2} -\hat{b}_{\bm{R}-\bm{a}_2}).\label{Kagomew}
\end{align}
One may also add a term
\begin{align}
\hat{\tilde{H}}^{(-)}&=\sum_{\bm{R}\in\Lambda}\tilde{\nu}\hat{w}_{\bm{R}}\hat{w}_{\bm{R}}^\dagger\label{Kagomeaddition}
\end{align}
that induces a negative dispersion for the flat band [Fig.~\ref{figKagome} (d)]:
\begin{align}
&\tilde{\omega}_{\bm{k}}^{(-)}= -\frac{\tilde{\nu}}{3}e_{\bm{k}}.
\end{align}

Even for general values of $\mu_1>0,\mu_2>0,\tilde{\nu}>0$, the lower two bands touch at $k_1=k_2=0$ and the form of the band touch is generally quadratic:
\begin{align}
\omega_{\bm{k}}^{(-)}=-\frac{\tilde{\nu}}{4}|\bm{k}|^2+O(|\bm{k}|^4),\quad \omega_{\bm{k}1}^{(+)}=\frac{\mu_2}{8}|\bm{k}|^2+O(|\bm{k}|^4).
\end{align}

\begin{figure}[t]
\begin{center}
\includegraphics[width=\columnwidth]{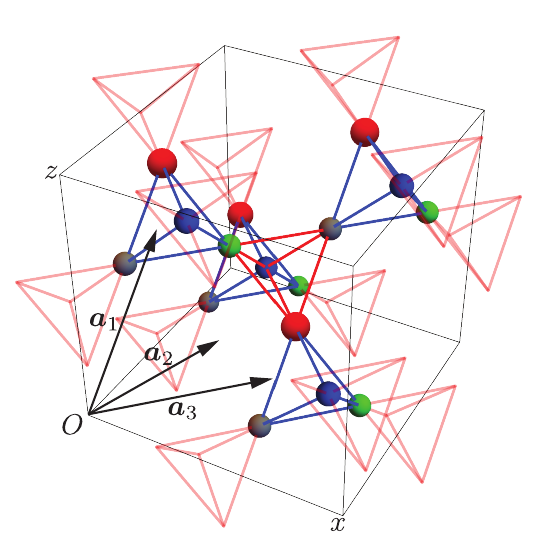}
\end{center}
\caption{The pyrochlore lattice.
\label{figPyrochlore}
}
\end{figure}

\subsection{Pyrochlore lattice}
The pyrochlore lattice is a 3D generalization of the Kagome lattice. The tight-biding model with nearest neighbor hopping is known to have a flat band~\cite{Hotta}.
Let us set $\bm{a}_1=(0,1,1)/2$, $\bm{a}_2=(1,0,1)/2$, $\bm{a}_3=(1,1,0)/2$.
Sublattices $\hat{a}_{\bm{R}}$, $\hat{b}_{\bm{R}}$, $\hat{c}_{\bm{R}}$, $\hat{d}_{\bm{R}}$ are placed at
$(5,3,3)/8$, $(3,5,3)/8$, $(3,3,5)/8$, $(5,5,5)/8$, respectively.

If the nearest-neighbor bopping is $t$ and the onsite potential is $\mu$, the Hamiltonian can be written as $\hat{H}=\sum_{\bm{R}\in\Lambda}\hat{\bm{c}}_{\bm{R}}^\dagger h\hat{\bm{c}}_{\bm{R}}$, where
$\hat{\bm{c}}_{\bm{R}}^\dagger=(\hat{a}_{\bm{R}}^\dagger,\hat{b}_{\bm{R}}^\dagger,\hat{c}_{\bm{R}}^\dagger,\hat{d}_{\bm{R}}^\dagger,\hat{a}_{\bm{R}+\bm{a}_1}^\dagger,\hat{b}_{\bm{R}+\bm{a}_2}^\dagger,\hat{c}_{\bm{R}+\bm{a}_3}^\dagger)$ and
\begin{align}
h=t\begin{pmatrix}
0&1&1&1&0&0&0\\
1&0&1&1&0&0&0\\
1&1&0&1&0&0&0\\
1&1&1&0&1&1&1\\
0&0&0&1&0&1&1\\
0&0&0&1&1&0&1\\
0&0&0&1&1&1&0
\end{pmatrix}+\frac{\mu}{2}\begin{pmatrix}
1&0&0&0&0&0&0\\
0&1&0&0&0&0&0\\
0&0&1&0&0&0&0\\
0&0&0&2&0&0&0\\
0&0&0&0&1&0&0\\
0&0&0&0&0&1&0\\
0&0&0&0&0&0&1
\end{pmatrix}.
\end{align}
When $\mu=2t$, we find $\hat{H}_{\bm{R}}=\hat{H}_{\bm{R}}^{(+)}=\sum_{\alpha=1,2}\mu_{\alpha}\hat{\psi}_{\bm{R}\alpha}^\dagger\hat{\psi}_{\bm{R}\alpha}$ with $\mu_{1}=5t$, $\mu_{2}=3t$ and
\begin{align}
\hat{\psi}_{\bm{R}1}&=\frac{\hat{a}_{\bm{R}}+\hat{b}_{\bm{R}}+\hat{c}_{\bm{R}}+2\hat{d}_{\bm{R}}+\hat{a}_{\bm{R}+\bm{a}_1}+\hat{b}_{\bm{R}+\bm{a}_2}+\hat{c}_{\bm{R}+\bm{a}_3}}{\sqrt{10}},\\
\hat{\psi}_{\bm{R}2}&=\frac{\hat{a}_{\bm{R}}+\hat{b}_{\bm{R}}+\hat{c}_{\bm{R}}-\hat{a}_{\bm{R}+\bm{a}_1}-\hat{b}_{\bm{R}+\bm{a}_2}-\hat{c}_{\bm{R}+\bm{a}_3}}{\sqrt{6}}.
\end{align}
The Hamiltonian in the momentum space is $\hat{H}=\sum_{\bm{k}}\hat{\bm{c}}_{\bm{k}}^\dagger H_{\bm{k}}\hat{\bm{c}}_{\bm{k}}$ with $\hat{\bm{c}}_{\bm{k}}^\dagger=(\hat{a}_{\bm{k}}^\dagger,\hat{b}_{\bm{k}}^\dagger,\hat{c}_{\bm{k}}^\dagger,\hat{d}_{\bm{k}}^\dagger)$ and
\begin{align}
H_{\bm{k}}=H_{\bm{k}}^{(+)}=t\begin{pmatrix}
2&1+e_{\bm{k}}^{21}&1+e_{\bm{k}}^{31}&1+e^{-ik_1}\\
1+e_{\bm{k}}^{12}&2&1+e_{\bm{k}}^{32}&1+e^{-ik_2}\\
1+e_{\bm{k}}^{13}&1+e_{\bm{k}}^{23}&2&1+e^{-ik_3}\\
1+e^{ik_1}&1+e^{ik_2}&1+e^{ik_3}&2\\
\end{pmatrix}.
\end{align}
Here, we introduced a shorthand $e_{\bm{k}}^{\ell m}\coloneqq e^{i(k_\ell -k_m)}$. 
The band dispersions are
\begin{align}
&\omega_{\bm{k}1}^{(+)}= 4t-t\sqrt{16-2e_{\bm{k}}},\\
&\omega_{\bm{k}2}^{(+)}= 4t+t\sqrt{16-2e_{\bm{k}}},\\
&\omega_{\bm{k}1}^{(0)}=\omega_{\bm{k}2}^{(0)}= 0
\end{align}
with $e_{\bm{k}}\coloneqq 6-\cos k_1-\cos k_2-\cos k_3-\cos (k_1-k_2)-\cos (k_2-k_3)-\cos (k_3-k_1)$.  The unnormalized Bloch states for the flat bands are
\begin{align}
&\bm{w}_{\bm{k}1}=\frac{1}{\sqrt{6}}\begin{pmatrix}
0\\
-1+e^{ik_3}\\
1-e^{ik_2}\\
e^{ik_2}-e^{ik_3}
\end{pmatrix},
\bm{w}_{\bm{k}2}=\frac{1}{\sqrt{6}}\begin{pmatrix}
1-e^{ik_3}\\
0\\
-1+e^{ik_1}\\
e^{ik_3}-e^{ik_1}
\end{pmatrix},
\end{align}
which corresponds to the hexagonal states 
\begin{align}
\hat{w}_{\bm{R}1}&=\frac{\hat{c}_{\bm{R}}-\hat{b}_{\bm{R}}+\hat{d}_{\bm{R}-\bm{a}_2}-\hat{c}_{\bm{R}-\bm{a}_2}+\hat{b}_{\bm{R}-\bm{a}_3}-\hat{d}_{\bm{R}-\bm{a}_3}}{\sqrt{6}},\\
\hat{w}_{\bm{R}2}&=\frac{\hat{a}_{\bm{R}}-\hat{c}_{\bm{R}}+\hat{d}_{\bm{R}-\bm{a}_3}-\hat{a}_{\bm{R}-\bm{a}_3}+\hat{c}_{\bm{R}-\bm{a}_1}-\hat{d}_{\bm{R}-\bm{a}_1}}{\sqrt{6}}.
\end{align}
The choice of the unnormalized Bloch states for the flat bands is not unique. For example, 
\begin{align}
&\bm{w}_{\bm{k}3}=\frac{1}{\sqrt{6}}\begin{pmatrix}
-1+e^{ik_2}\\
1-e^{ik_1}\\
0\\
e^{ik_1}-e^{ik_2}
\end{pmatrix},
\bm{w}_{\bm{k}4}=\frac{1}{\sqrt{6}}\begin{pmatrix}
e^{ik_2}-e^{ik_3}\\
e^{ik_3}-e^{ik_1}\\
e^{ik_1}-e^{ik_2}\\
0
\end{pmatrix}
\end{align}
also correspond to hexagonal states:
\begin{align}
\hat{w}_{\bm{R}3}&=\frac{\hat{b}_{\bm{R}}-\hat{a}_{\bm{R}}+\hat{d}_{\bm{R}-\bm{a}_1}-\hat{b}_{\bm{R}-\bm{a}_1}+\hat{a}_{\bm{R}-\bm{a}_2}-\hat{d}_{\bm{R}-\bm{a}_2}}{\sqrt{6}},\\
\hat{w}_{\bm{R}4}&=\frac{\hat{a}_{\bm{R}-\bm{a}_2}-\hat{b}_{\bm{R}-\bm{a}_1}+\hat{c}_{\bm{R}-\bm{a}_1}-\hat{a}_{\bm{R}-\bm{a}_3}+\hat{b}_{\bm{R}-\bm{a}_3}-\hat{c}_{\bm{R}-\bm{a}_2}}{\sqrt{6}}.
\end{align}
We find that if we use all of these four hexagonal states by adding
\begin{align}
\hat{\tilde{H}}^{(-)}=\sum_{\bm{R}\in\Lambda}\sum_{\ell=1}^4\tilde{\nu}\hat{w}_{\bm{R}\ell}\hat{w}_{\bm{R}\ell}^\dagger
\end{align}
to the Hamiltonian, we can induce a dispersion 
\begin{align}
&\tilde{\omega}_{\bm{k}1}^{(-)}=\tilde{\omega}_{\bm{k}2}^{(-)}=-\frac{\tilde{\nu}}{3}e_{\bm{k}}
\end{align}
to the flat bands without breaking the symmetry of the pyrochlore lattice.

\section{U(1) breaking models}
\label{sec:BdG}
So far, we have considered free fermions with U(1) symmetry.
In this section, we discuss the frustration-free condition without U(1) symmetry.

\subsection{Derivation of BdG form}
Let us first show that the most general local Hamiltonian can be expressed in the Bogoliubov-de Gennes (BdG) form: $\hat{H} = \sum_{\bm{R} \in \Lambda} \hat{H}_{\bm{R}}$ with 
\begin{align}
	\hat{H}_{\bm{R}} = \hat{\Psi}_{\bm{R}}^\dagger h_{\bR}\hat{\Psi}_{\bm{R}} + C_{\bm{R}},
\end{align}
where $\hat{\Psi}_{\bm{R}}^\dagger \coloneqq \begin{pmatrix}
	\hat{\bm{c}}_{\bm{R}}^\dagger& \hat{\bm{c}}_{\bm{R}}^\top 
\end{pmatrix}$
is the Nambu spinor and $h_{\bR}$ is a $2N_r$-dimensional matrix with the particle-hole symmetry
\begin{align}
U_{\mathcal{P}}h_{\bR}^{\top}U_{\mathcal{P}}^\dagger = -h_{\bR},\quad U_{\mathcal{P}}\coloneqq
\begin{pmatrix}
O & \mathbbm{1} \\
\mathbbm{1} & O
\end{pmatrix}.\label{eq:PHS}
\end{align}
The Nambu spinor contains $N_r$ pairs of creation and annihilation operators and satisfies $\hat{\Psi}_{\bm{R}}^\dagger = \hat{\Psi}_{\bm{R}}^{\top}U_{\mathcal{P}}$.

To derive the BdG form, let us start with a general Hamiltonian on the Majorana basis:
\begin{align}
	\label{eq:BdG_HR}
	\hat{H}_{\bm{R}} \coloneqq i\hat{\bm{\gamma}}_{\bm{R}}^{\top}A_{\bR}\hat{\bm{\gamma}}_{\bm{R}} + C_{\bR},
\end{align}
where $A_{\bR}$ is a  $2N_r$-dimensional real antisymmetric matrix and $\hat{\bm{\gamma}}_{\bm{R}}$ is a  vector composed of $2N_r$ Majorana fermions $\hat{\gamma}_{\bm{R}'\tau i}$ within the range $r$ from $\bm{R}$.
The Majorana operators satisfy $\hat{\gamma}_{\bm{R}\tau i}^\dagger = \hat{\gamma}_{\bm{R}\tau i}$ $(i=1,2)$ and $\{\hat{\gamma}_{\bm{R}\tau i}, \hat{\gamma}_{\bm{R}'\tau' j}\} = 2 \delta_{\bm{R}\bm{R}'}\delta_{\tau\tau'}\delta_{ij}$.
We define complex fermions by 
\begin{align}
	&\hat{c}_{\bm{R}\tau} \coloneqq \frac{1}{2}\left(\hat{\gamma}_{\bm{R}\tau1} + i \hat{\gamma}_{\bm{R}\tau2}\right).
\end{align}
Then the vector $\hat{\Psi}_{\bm{R}}$ and the matrix $h_{\bR}$ in the BdG form are given by
\begin{align}
	&\hat{\Psi}_{\bm{R}} = \frac{1}{2}\begin{pmatrix}
		\mathbbm{1} & i \mathbbm{1} \\
		\mathbbm{1} & -i \mathbbm{1}
	\end{pmatrix}\hat{\bm{\gamma}}_{\bm{R}},\\
	&h_{\bR} = \begin{pmatrix}
		\mathbbm{1} & i \mathbbm{1} \\
		\mathbbm{1} & -i \mathbbm{1}
	\end{pmatrix}
	i A_{\bR}
	\begin{pmatrix}
		\mathbbm{1} & \mathbbm{1} \\
		-i \mathbbm{1} & i\mathbbm{1}
	\end{pmatrix}.
\end{align}
For example, the symmetry condition~\eqref{eq:PHS} can be shown as
\begin{align}
	&U_{\mathcal{P}}h_{\bR}^\top U_{\mathcal{P}}^\dagger \nonumber \\
	&=
	\begin{pmatrix}
		O & \mathbbm{1} \\
		\mathbbm{1} & O
	\end{pmatrix}
	\begin{pmatrix}
		\mathbbm{1} & -i\mathbbm{1} \\
		\mathbbm{1} & i \mathbbm{1}
	\end{pmatrix}i A_{\bR}^\top
	\begin{pmatrix}
		\mathbbm{1} & \mathbbm{1} \\
		i \mathbbm{1} & -i \mathbbm{1}
	\end{pmatrix}
	\begin{pmatrix}
		O & \mathbbm{1} \\
		\mathbbm{1} & O
	\end{pmatrix}\nonumber \\
	&= 
	\begin{pmatrix}
		\mathbbm{1} & i \mathbbm{1} \\
		\mathbbm{1} & -i \mathbbm{1}
	\end{pmatrix}
	(-i A_{\bR})
	\begin{pmatrix}
		\mathbbm{1} & \mathbbm{1} \\
		-i \mathbbm{1} & i\mathbbm{1}
	\end{pmatrix} = -h_{\bR}.
\end{align}

\subsection{Frustration-free condition}
Now, we are ready to derive the frustration-free condition. Let 
$\bm{\psi}_{\bm{R}\alpha}$ $(\alpha=1,2,\cdots,A_{\bm{R}})$ be eigenvectors of $h_{\bR}$ with positive eigenvalues $\mu_{\bm{R}\alpha}/2>0$. 
Due to the symmetry relation \eqref{eq:PHS}, $\bm{\phi}_{\bm{R}\alpha}\coloneqq U_{\mathcal{P}}\bm{\psi}^{*}_{\bm{R}\alpha}$ is an eigenvector with a negative eigenvalue $-\mu_{\bm{R}\alpha}/2<0$; that is, 
\begin{align}
h_{\bm{R}}&=\sum_{\alpha=1}^{A_{\bm{R}}}\frac{\mu_{\bm{R}\alpha}}{2}\bm{\psi}_{\bm{R}\alpha}\bm{\psi}_{\bm{R}\alpha}^\dagger-\sum_{\alpha=1}^{A_{\bm{R}}}\frac{\mu_{\bm{R}\alpha}}{2}\bm{\psi}_{\bm{R}\alpha}^*\bm{\psi}_{\bm{R}\alpha}^\top.\label{HRBDG}
\end{align}
We define corresponding fermionic operators by
\begin{align}
\hat{\psi}_{\bm{R}\alpha}= \bm{\psi}_{\bm{R}\alpha}^\dagger\hat{\Psi}_{\bm{R}}.
\end{align}
In this case, $\hat{\phi}_{\bm{R}\alpha}=\bm{\phi}_{\bm{R}\alpha}^\dagger \hat{\Psi}_{\bm{R}}$ coincides with $\hat{\psi}_{\bm{R}\alpha}^\dagger$ and we will not use it in the following. The anticommutation relations 
\begin{align}
\{\hat{\psi}_{\bm{R}\alpha}, \hat{\psi}_{\bm{R}\alpha'}^\dagger\} =\bm{\psi}_{\bm{R}\alpha}^\dagger\bm{\psi}_{\bm{R}\alpha'}= \delta_{\alpha\alpha'}\label{AL1}
\end{align}
and $\{\hat{\psi}_{\bm{R}\alpha}, \hat{\psi}_{\bm{R}\alpha}\}  =\bm{\psi}_{\bm{R}\alpha}^\dagger\bm{\phi}_{\bm{R}\alpha}= 0$ follow from the anticommutation relations among $\hat{c}_{\bm{R}\tau}$ and the orthogonality of eigenvectors of $h_{\bm{R}}$. 

Choosing $C_{\bm{R}}=\sum_{\alpha=1}^{A_{\bm{R}}}\mu_{\bm{R}\alpha}/2$, we rewrite $\hat{H}_{\bm{R}}$ as
\begin{align}
	\hat{H}_{\bm{R}} &= \sum_{\alpha=1}^{A_{\bm{R}}}\frac{\mu_{\bm{R}\alpha}}{2}\big( \hat{\psi}_{\bm{R}\alpha}^\dagger\hat{\psi}_{\bm{R}\alpha}+ \hat{\phi}_{\bm{R}\alpha}\hat{\phi}_{\bm{R}\alpha}^\dagger\big)\notag\\
	&=\sum_{\alpha=1}^{A_{\bm{R}}}\mu_{\bm{R}\alpha} \hat{\psi}_{\bm{R}\alpha}^\dagger\hat{\psi}_{\bm{R}\alpha}.
	\label{BdG2}
\end{align}
Consider a state given by $\ket{\Phi_{\bm{R}}} \coloneqq \prod_{\alpha=1}^{A_{\bm{R}}}\hat{\psi}_{\bm{R}\alpha}\ket{0}$, which satisfies $\hat{\psi}_{\bm{R}\alpha}\ket{\Phi_{\bm{R}}} = 0$ for all  $\alpha$.
Hence, $\ket{\Phi_{\bm{R}}} $ is a ground state of $\hat{H}_{\bm{R}}$ and satisfies $\hat{H}_{\bm{R}}\ket{\Phi_{\bm{R}}} = 0$. 
Following the same discussion as the U(1)-symmetric case, we arrive at the frustration-free condition
\begin{align}
\label{eq:BdG_conditiontobeFFFF}
\{\hat{\psi}_{\bm{R}\alpha},\hat{\psi}_{\bm{R}'\alpha'}\}=0\quad\text{for all $\bm{R}$, $\bm{R}'$, $\alpha$, $\alpha'$.}
\end{align}

The Hamiltonian 
\begin{align}
\hat{H}=\sum_{\bm{R}\in\Lambda}\sum_{\alpha=1}^{A_{\bm{R}}}\mu_{\bm{R}\alpha} \hat{\psi}_{\bm{R}\alpha}^\dagger\hat{\psi}_{\bm{R}\alpha}\label{BdG3}
\end{align}
with anticommutation relation in \eqref{AL1} and \eqref{eq:BdG_conditiontobeFFFF} can be accurately represented by an auxiliary U(1) symmetric Hamiltonian in which the number of fermion species is doubled. To see this correspondence concretely, let us introduce $\hat{f}_{\bm{R}\tau i}$ ($i=1,2$) for every $\hat{c}_{\bm{R}\tau}$.
These operators are complex fermions satisfying 
$\{\hat{f}_{\bm{R}\tau i},\hat{f}_{\bm{R}'\tau'j}\}=0$ and $\{\hat{f}_{\bm{R}\tau i},\hat{f}_{\bm{R}'\tau'j}^\dagger\}=\delta_{\bm{R},\bm{R}'}\delta_{\tau,\tau'}\delta_{i,j}$. 
We replace the Nambu spinor $\hat{\Psi}_{\bm{R}}^\dagger \coloneqq \begin{pmatrix}\hat{\bm{c}}_{\bm{R}}^\dagger& \hat{\bm{c}}_{\bm{R}}^\top \end{pmatrix}$ by
$\hat{\tilde{\Psi}}_{\bm{R}}^\dagger \coloneqq \begin{pmatrix}\hat{\bm{f}}_{\bm{R},1}^\dagger& \hat{\bm{f}}_{\bm{R},2}^\dagger \end{pmatrix}$ and $h_{\bm{R}}$  in Eq.~\eqref{HRBDG} by 
\begin{align}
\tilde{H}_{\bm{R}}\coloneqq\sum_{\alpha=1}^{A_{\bm{R}}}\mu_{\bm{R}\alpha}\bm{\psi}_{\bm{R}\alpha}\bm{\psi}_{\bm{R}\alpha}^\dagger.
\end{align}
The operator $\hat{\tilde{\psi}}_{\bm{R}\alpha}\coloneqq \bm{\psi}_{\bm{R}\alpha}^\dagger\hat{\tilde{\Psi}}_{\bm{R}}$
satisfies not only 
\begin{align}
\{\hat{\tilde{\psi}}_{\bm{R}\alpha},\hat{\tilde{\psi}}_{\bm{R}\alpha'}^\dagger\}=\delta_{\alpha,\alpha'}\label{AL2}
\end{align}
within a unit cell $\bm{R}$ but also
\begin{align}
\{\hat{\tilde{\psi}}_{\bm{R}\alpha},\hat{\tilde{\psi}}_{\bm{R}'\alpha'}\}=0\quad \text{for all $\alpha,\alpha', \bm{R}, \bm{R}'$}\label{eq:BdG_conditiontobeFFFF2}
\end{align}
as a trivial relation for the U(1) symmetric case. Observe that the anticommutation relations of $\hat{\tilde{\psi}}_{\bm{R}\alpha}$'s in \eqref{AL2} and \eqref{eq:BdG_conditiontobeFFFF2} and $\hat{\psi}_{\bm{R}\alpha}$'s in \eqref{AL1} and \eqref{eq:BdG_conditiontobeFFFF} are the same.
As a consequence, the spectrum of 
\begin{align}
\hat{\tilde{H}}&\coloneqq\sum_{\bm{R}\in\Lambda}\hat{\tilde{\Psi}}_{\bm{R}}^\dagger \tilde{h}_{\bm{R}}\hat{\tilde{\Psi}}_{\bm{R}}=\sum_{\bm{R}\in\Lambda}\sum_{\alpha=1}^{A_{\bm{R}}}\mu_{\bm{R}\alpha} \hat{\tilde{\psi}}_{\bm{R}\alpha}^\dagger\hat{\tilde{\psi}}_{\bm{R}\alpha}
\end{align}
and $\hat{H}$ in \eqref{BdG3} is identical, except for the degeneracy originating from the flat bands associated with $\hat{\tilde{\phi}}_{\bm{R}\alpha}\coloneqq\bm{\phi}_{\bm{R}\alpha}^\dagger \hat{\tilde{\Psi}}_{\bm{R}}$. 
In particular, if the system is gapless then the low-energy dispersion is quadratic or softer as shown in Sec.~\ref{sec:U1}.  In Appendix~\ref{app:BdG}, we present an alternative derivation which does not invoke the auxiliary fermionic system.

\subsection{Example}
As an example, let us discuss a fermionic model that corresponds to the combination of the XY spin chain under a magnetic field and the cluster model:
\begin{align}
\hat{H}=\sum_{i=1}^L\Big(J_x\hat{\sigma}_i^x\hat{\sigma}_{i+1}^x+J_y\hat{\sigma}_i^y\hat{\sigma}_{i+1}^y+B\hat{\sigma}_i^z-A\hat{\sigma}_{i-1}^x\hat{\sigma}_i^z\hat{\sigma}_{i+1}^x\Big).
\end{align}
Introducing the Jordan-Wigner transformation
\begin{align}
\hat{\sigma}_i^x&=e^{i\pi\sum_{k=1}^{i-1}\hat{c}_j^\dagger \hat{c}_j}(\hat{c}_i^\dagger+\hat{c}_i),\\
\hat{\sigma}_i^y&=-ie^{i\pi\sum_{k=1}^{i-1}\hat{c}_j^\dagger \hat{c}_j}(\hat{c}_i^\dagger-\hat{c}_i),\\
\hat{\sigma}_i^z&=2\hat{c}_i^\dagger \hat{c}_i-1,
\end{align}
we obtain the fermionic representation
\begin{align}
\hat{H}&=\sum_{i=1}^L\Big(-t(\hat{c}_i^\dagger\hat{c}_{i+1}+\hat{c}_{i+1}^\dagger\hat{c}_i)-\Delta(\hat{c}_{i+1}\hat{c}_i+\hat{c}_i^\dagger\hat{c}_{i+1}^\dagger)\notag\\
&+\mu(\hat{c}_i^\dagger \hat{c}_i-1/2)+\frac{1}{2}t'(\hat{c}_{i-1}^\dagger-\hat{c}_{i-1})(\hat{c}_{i+1}^\dagger+\hat{c}_{i+1})\Big),
\end{align}
where $t=-(J_x+J_y)$, $\Delta=J_y-J_x$, $\mu=2B$, and $t'=2A$. We assume the periodic boundary condition for the fermionic Hamiltonian.

\subsubsection{The Kitaev chain}
\label{sec:kitaev}
Let us start with the case $t'=0$. The model is equivalent to the Kitaev chain~\cite{kitaev_chain}, whose local Hamiltonian can be written as
\begin{align}
	\label{eq:Hamiltonian_kitaev}
	\hat{H}_i\coloneqq \frac{t_0}{2}+\frac{1}{2}\hat{\Psi}_i^\dagger
	\begin{pmatrix}
		\frac{\mu}{2} & -t & 0 & -\Delta \\
		-t & \frac{\mu}{2} & \Delta & 0 \\
		0 & \Delta & -\frac{\mu}{2} & t \\
		-\Delta & 0 & t & -\frac{\mu}{2} \\
	\end{pmatrix}\hat{\Psi}_i
\end{align}
with $\hat{\Psi}_i^\top \coloneqq \begin{pmatrix}\hat{c}_i & \hat{c}_{i+1} & \hat{c}_i^\dagger &\hat{c}_{i+1}^\dagger\end{pmatrix}$. When $t=t_0/2>0$, $\mu=t_0\cos2\theta$, and $\Delta=-t_0\sin\theta\cos\theta$, the Hamiltonian is frustration-free regardless of the choice of  $\theta$~\cite{PhysRevB.92.115137}\footnote{In Ref.~\cite{PhysRevB.92.115137}, a nearest-neighbor interaction $U$ is added while maintaining the frustration-free property of the model.}. Indeed, we find $\hat{H}_i=t_0\hat{\psi}_i^\dagger \hat{\psi}_i$ with 
\begin{align}
\hat{\psi}_i=\frac{\cos\theta}{\sqrt{2}}(\hat{c}_i-\hat{c}_{i+1})+\frac{\sin\theta}{\sqrt{2}}(\hat{c}_i^\dagger+\hat{c}_{i+1}^\dagger)
\end{align}
and $\{\hat{\psi}_i,\hat{\psi}_j\}=0$ for any $i,j$. 

If we define the operator
\begin{align}
\hat{\alpha}_k&\coloneqq \frac{1}{\sqrt{L}}\sum_{j=1}^Le^{-ikj}\hat{\psi}_j\notag\\
&=\frac{\cos\theta}{\sqrt{2}}(1-e^{ik})\hat{c}_k+\frac{\sin\theta}{\sqrt{2}}(1+e^{ik})\hat{c}_{-k}^\dagger,
\end{align}
which satisfies $\{\hat{\alpha}_k,\hat{\alpha}_{k'}^\dagger\}=\delta_{k,k'}(1-\cos 2\theta\cos k)$, the Hamiltonian is written as $\hat{H}=\sum_kt_0\hat{\alpha}_k^\dagger \hat{\alpha}_k$. Therefore, the low-energy dispersion is given by $\omega_k=t_0(1-\cos2\theta\cos k)$. Unless  $\theta\neq0$ or $\pi/2$, the excitation gap originates from $\Delta\neq0$. When $\theta=0$ or $\pi/2$, $\Delta$ vanishes and $\omega_k$ is gapless and quadratic~\cite{PhysRevB.98.155119}.

This dispersion can also be derived from the standard method. After the Fourier transformation, the BdG Hamiltonian can be expressed as
\begin{align}
\label{eq:k_kitaev}
\hat{H}&=\frac{Lt_0}{2}+\sum_k\begin{pmatrix}
\hat{c}^\dagger_{k}&\hat{c}_{-k}
\end{pmatrix}(h_k\sigma_3+\delta_k\sigma_2)
\begin{pmatrix}
\hat{c}_{k}\\\hat{c}^\dagger_{-k}
\end{pmatrix}\notag\\
&=\sum_k\omega_k\hat{\gamma}_k^\dagger \hat{\gamma}_k,
\end{align}
where $h_k=\tfrac{1}{2}\mu-t\cos k$ and $\delta_k=\Delta \sin k$, and $\omega_k=2\sqrt{h_k^2+\delta_k^2}=t_0(1-\cos2\theta\cos k)$.

\subsubsection{Kinetic Ising model}
Next, let us consider the case $t'\neq0$.
\begin{align}
\hat{H}_i
&=t_0+\frac{1}{4}
\hat{\Psi}_i^\dagger
\begin{pmatrix}
0&-t&t'&0&-\Delta&t'\\
-t&2\mu&-t&\Delta&0&-\Delta\\
t'&-t&0&-t'&\Delta&0\\
0&\Delta&-t'&0&t&-t'\\
-\Delta&0&\Delta&t&-2\mu&t\\
t'&-\Delta&0&-t'&t&0
\end{pmatrix}\hat{\Psi}_i,
\end{align}
where 
$\hat{\Psi}_i=\begin{pmatrix}
\hat{c}_{i-1}&
\hat{c}_i&
\hat{c}_{i+1}&
\hat{c}_{i-1}^\dagger&
\hat{c}_i^\dagger&
\hat{c}_{i+1}^\dagger
\end{pmatrix}^{\top}$.
When $t=\Delta=t_0\sin\theta\cos\theta$ ($t_0>0$), $\mu=t_0\cos^2\theta$, and $t'=t_0\sin^2\theta$, the Hamiltonian is frustration-free. We find $\hat{H}_i=t_0\hat{\psi}_i^\dagger \hat{\psi}_i$ with
\begin{align}
\hat{\psi}_i=\frac{1}{2}\cos\theta(\hat{c}_{i-1}+\hat{c}_{i+1}-\hat{c}_{i-1}^\dagger+\hat{c}_{i+1}^\dagger)-\sin\theta\,\hat{c}_i.
\end{align}
Since $\{\hat{\psi}_i,\hat{\psi}_j\}=0$ for any $i,j$, the Hamiltonian is frustration free.

The operator
\begin{align}
\hat{\alpha}_k&\coloneqq \frac{1}{\sqrt{L}}\sum_{j=1}^Le^{-ikj}\hat{\psi}_j\notag\\
&=\cos\theta(\cos k\hat{c}_k+i\sin k\hat{c}_{-k}^\dagger)-\sin\theta \hat{c}_{k}
\end{align}
satisfies $\{\hat{\alpha}_k,\hat{\alpha}_{k'}^\dagger\}=\delta_{k,k'}(1-\sin 2\theta\cos k)$ and  the Hamiltonian becomes $\hat{H}=\sum_kt_0\hat{\alpha}_k^\dagger \hat{\alpha}_k$. Hence, the dispersion is given by $\omega_k=t_0(1-\sin2\theta\cos k)$.

This model can be mapped to the kinetic Ising model by replacing $\hat{\sigma}_i^x\to \hat{\sigma}_i^z$ and $\hat{\sigma}_i^z\to -\hat{\sigma}_i^x$ and setting $2\beta J=\mathrm{arctanh}(\sin2\theta)$:
\begin{align}
\hat{H}_i=\frac{t_0}{2\cosh(\beta J(\hat{\sigma}_{i-1}^z+\hat{\sigma}_{i+1}^z))}\Big(e^{-\beta J\hat{\sigma}_i^z(\hat{\sigma}_{i-1}^z+\hat{\sigma}_{i+1}^z)}-\hat{\sigma}_i^x\Big).
\end{align}
At $\theta=\pi/4$, which corresponds to $\beta J=\infty$ (i.e., the zero temperature), the model reduces to the uncle Hamiltonian for the GHZ state:
\begin{align}
\hat{H}_i=\frac{t_0}{4}\big(2-\hat{\sigma}_{i-1}^z\hat{\sigma}_{i}^z-\hat{\sigma}_i^z\hat{\sigma}_{i+1}^z-\hat{\sigma}_i^x+\hat{\sigma}_{i-1}^z\hat{\sigma}_i^x\hat{\sigma}_{i+1}^z\big).
\end{align}
The gapless excitation of the model is associated with the critical nature of the classical Ising model at zero temperature.

	\section{Topological bands in frustration-free free-fermionic models}
	\label{sec:topologicalphases}
	In this section, we investigate what kinds of topological phases can be realized by frustration-free free fermionic Hamiltonians.
	
	\subsection{Translationally invariance and Wannier localizability}
	We consider the Hamiltonian in momentum space:
	\begin{align}
		\hat{H} = \sum_{\bm{R} \in \Lambda} \hat{H}_{\bm{R}} =\sum_{\bm{k}} \hat{\bm{c}}_{\bm{k}}^\dagger H_{\bm{k}} \hat{\bm{c}}_{\bm{k}}.
	\end{align}
	The matrix $H_{\bm{k}}$ can always be decomposed into two Hermitian matrices $H^{(+)}_{\bm{k}}$ and $H^{(-)}_{\bm{k}}$:
	\begin{align}
		H_{\bm{k}} &= H^{(+)}_{\bm{k}} + H^{(-)}_{\bm{k}},
	\end{align}
	whose eigenvalues are denoted by $\omega_{n\bm{k}}^{(+)} \geq 0\ (n=1,2,\cdots,N^{(+)})$ and $\omega_{m\bm{k}}^{(-)} \leq 0\ (m=1,2,\cdots,N^{(-)})$, respectively.
	
	We prove the following theorem: Consider a translationally invariant free fermionic Hamiltonian $\hat{H}$ with the corresponding Hermitian matrix decomposition $H_{\bm{k}} = H^{(+)}_{\bm{k}} + H^{(-)}_{\bm{k}}$.
	If $\hat{H}$ is frustration-free and $\omega_{m\bm{k}}^{(-)} < 0$ for all $m$ and $\bm{k}$, there exists an (over)complete set of compactly supported Wannier-type functions.
	
	What we have to show is as follows.
	Let us define the space of occupied states by
	\begin{align}
		\mathcal{E}_{\text{occ}} \coloneqq \mathrm{Span}_{n, \bm{k}}\left\{\ket{\gamma_{n\bm{k}}}: \hat{H}\ket{\gamma_{n\bm{k}}} = \epsilon_{n\bm{k}}\ket{\gamma_{n\bm{k}}}, \epsilon_{n\bm{k}} < 0\right\},
	\end{align}
	where $\hat{H} = \sum_{\bm{k}, n} \epsilon_{n\bm{k}}\hat{\gamma}_{n\bm{k}}^\dagger \hat{\gamma}_{n\bm{k}}$ and $|\gamma_{n\bm{k}}\rangle = \hat{\gamma}_{n\bm{k}}^\dagger\ket{0}$.
	It suffices to show that there exists an (over)complete set of orbitals $\{\ket{f_{n}}\}_{n}$ such that they span $\mathcal{E}_{\text{occ}}$ and their supports are strictly finite.
	We claim that $\{\ket{\phi_{\bm{R}\beta}} \coloneqq \hat{\phi}^{\dagger}_{\bm{R}\beta}\ket{0}\}_{R, \beta}$ is indeed such a set.
	
	To prove this claim, we use the following basic knowledge of free-fermionic systems:
	Let $\{|f_n \rangle \coloneqq \hat{f}^{\dagger}_{n}\ket{0}\}_{n=1}^{M}$ be a set of mutually orthonormal one-particle states.
	Define the space $\mathcal{W}_{f} \coloneqq \mathrm{Span} \left\{|f_n \rangle \right\}_{n=1}^{M}$ and the orthogonal projector $P_{\mathcal{W}_f} \coloneqq \sum_{n=1}^{M} |f_n \rangle \langle f_n |$.
	Suppose that a ground state $|\Phi\rangle$ is the Slater determinant of the $f$-orbitals, i.e.,
	\begin{equation}
		|\Phi\rangle = \hat{f}^{\dagger}_{1} \hat{f}^{\dagger}_{2} \cdots \hat{f}^{\dagger}_{M} \ket{0}.
	\end{equation}
	Then, the projector is equivalent to the two-point correlation function:
	\begin{equation}
		\label{eq:projector_relation}
		\begin{aligned}
			[P_{\mathcal{W}_f}]_{ij} = \langle i| P_{\mathcal{W}_f} |j \rangle = \langle \Phi | \hat{c}^{\dagger}_{j} \hat{c}_{i} | \Phi \rangle,
		\end{aligned}
	\end{equation}
	where $\{|i\rangle \coloneqq \hat{c}^{\dagger}_i | 0 \rangle\}_{i}$ is a basis set of the one-particle Hilbert space.
		For completeness, we provide the derivation of the relation~\eqref{eq:projector_relation} in Appendix~\ref{app:proof_lemma_projector}.
	
	Now, we are ready to prove the existence of the overcomplete set of compactly supported Wannier-type functions.
	Since we are interested in a gapped free fermionic Hamiltonian, the ground state is unique and is constructed by filling the orbitals defined by $\{\hat{\phi}^{\dagger}_{\bm{R}\beta}\}_{\bm{R}, \beta}$.
	We can always construct an orthonormal set of orbitals $\{|f_n \rangle\}_{n=1}^{VN^{(-)}}$ from linear combinations of $\{|\phi_{\bm{R}\beta}\rangle \}_{\bm{R}, \beta}$, and the ground state is written as the Slater determinant of these $f$-orbitals.
	Alternatively, the same ground state can be expressed as the Slater determinant of $\{\gamma^{\dagger}_{n\bm{k}}\}_{n, \bm{k}}$.
	Consequently, the projector onto $\mathrm{Span} \{|\phi_{\bm{R}\beta}\rangle \}_{\bm{R}, \beta}$ and that onto $\mathcal{E}_{\text{occ}}$ coincide.
	Thus, the spaces are identical.
	We conclude that $\{|\phi_{\bm{R}\beta}\rangle \}_{\bm{R}, \beta}$ is an (over)complete set of orbitals spanning $\mathcal{E}_{\text{occ}}$.
	By definition, $\hat{\phi}^{\dagger}_{\bm{R}\beta}$ has strictly local support around $\bm{R}$.
	This completes the proof.
	
	An analogous argument holds for the case where $\omega_{n\bm{k}}^{(+)} > 0$ for all $n$ and $\bm{k}$.
	In this case, $\{\ket{\psi_{\bm{R}\alpha}} \coloneqq \hat{\psi}^{\dagger}_{\bm{R}\alpha}\ket{0}\}_{\bm{R}, \alpha}$ constitutes an overcomplete set of compactly supported Wannier-type orbitals that span $\mathcal{E}_{\text{unocc}}$, defined by
	\begin{align}
		\mathcal{E}_{\text{unocc}} \coloneqq \mathrm{Span}_{n, \bm{k}}\left\{\ket{\gamma_{n\bm{k}}}: \hat{H}\ket{\gamma_{n\bm{k}}} = \epsilon_{n\bm{k}}\ket{\gamma_{n\bm{k}}}, \epsilon_{n\bm{k}} > 0\right\}.
	\end{align}
	The proof proceeds exactly as in the case where $\omega_{m\bm{k}}^{(-)} < 0$ for all $m$ and $\bm{k}$.
	
	It is natural to ask what kinds of topological phases allow for overcomplete sets of compactly supported Wannier-type functions.
	Ref.~\cite{PhysRevB.95.115309} addresses this question, classifying topological phases that admit such Wannier-type functions. It shows that they exist only in topological phases originating from the stacking of zero- and one-dimensional topological phases in the Altland-Zirnbauer symmetry classes.
	
	We present our argument on frustration-free realizable topological phases by combining free fermionic frustration-freeness and the results of Ref.~\cite{PhysRevB.95.115309}.
	If either $\omega_{m\bm{k}}^{(-)} < 0$ or $\omega_{n\bm{k}}^{(+)} > 0$ holds, then $H^{(-)}_{\bm{k}}$ or $H^{(+)}_{\bm{k}}$ admits an overcomplete set of compactly supported Wannier-type functions.
	As a result, the allowed stable topological phases are limited to the stacking of zero- and one-dimensional topological phases, as shown in Ref.~\cite{PhysRevB.95.115309}.
	Since a system must be topologically trivial when all bands are occupied, $H_{\bm{k}}^{(\pm)}$ must possess opposite topological invariants.
	Thus, the topological restriction on one of $H^{(\pm)}_{\bm{k}}$ applies to the other as well.
	We arrive at the following theorem:
	For translationally invariant frustration-free free fermionic Hamiltonians, if $\omega_{m\bm{k}}^{(-)} < 0$ or $\omega_{n\bm{k}}^{(+)} > 0$, stable topological phases arise from stacking  of zero- and one-dimensional topological phases.
	This theorem implies that other gapped topological phases, such as Chern insulators and $\mathbb{Z}_2$ topological insulators, cannot be realized by frustration-free free fermionic Hamiltonians.
	
	We make three remarks.
	First, we emphasize that strict locality is essential for the preceding argument.
	Various topological phases, such as those with nontrivial three-dimensional winding numbers in class AIII, are known to be Wannierizable in the sense that they admit \textit{exponentially} localized Wannier functions~\cite{Ono-Po-Watanabe2020, Wannierizability_Luka,Wannierizability_Nakamura,Wannierizability_Shiozaki}.
	However, such phases cannot be realized by frustration-free free fermionic Hamiltonians due to the lack of compactly supported Wannier-type functions.
	Second, we cannot rule out nontrivial topology beyond Ref.~\cite{PhysRevB.95.115309} if there exist two momenta $\bm{k}_1$ and $\bm{k}_2\ (\bm{k}_1 \neq \bm{k}_2)$ such that $\omega_{m\bm{k}_1}^{(-)} = 0$ and $\omega_{n\bm{k}_2}^{(+)} = 0$. 
	While the system is gapless, we can still discuss band topology of $H_{\bm{k}}^{(\pm)}$ even in this case.
	However, our argument does not apply immediately to such a system.
	Finally, we do not consider crystalline symmetries beyond lattice translation symmetry.
	Recent studies demonstrate that topological crystalline phases can be constructed from lower-dimensional topological phases protected by onsite symmetries~\cite{TC_PRB, TC_PRX, TC_AII, R-AHSS_Song, R-AHSS_Shiozaki,defect_network, Shiozaki-Ono2023, Ono-Shiozaki-Watanabe2022, WC_Fang}.
	Consequently, topological crystalline phases arising from decoupled wire constructions of one-dimensional topological phases can be realized by frustration-free free fermionic Hamiltonians, provided that each building block can be frustration-free. Furthermore, we conjecture that even in the presence of crystalline symmetries, frustration-free realizable free fermionic topological phases are limited to obstructed atomic limits~\cite{TQC,PhysRevResearch.3.013239} and decoupled wire constructions of one-dimensional topological phases.
	
	\subsection{Examples}
	We examine several examples to illustrate the validity of the general statement presented above.
		In Sec.~\ref{sec:F4_momentum_space}, we show that if a given free fermionic Hamiltonian is frustration-free, then $H_{\bm{k}}^{(\pm)}$ are Laurent polynomials of $e^{\ii \bm{k} \cdot \bm{a}_{i}}\ (i=1,2,\cdots,d)$.
		By contraposition, if $H_{\bm{k}}^{(\pm)}$ are not Laurent polynomials, then the Hamiltonian cannot be frustration-free.
	We utilize this statement to show that a given free fermionic Hamiltonian is not frustration-free.

	\subsubsection{Kitaev chain}
	We begin with the Kitaev chain introduced in Sec.~\ref{sec:kitaev}, whose Hamiltonian in momentum space is given in Eq.~\eqref{eq:k_kitaev}.
	The Hermitian matrix $H_{k}$ and the energy dispersion $\epsilon_{k}$ are given by
	\begin{equation}
		\begin{aligned}
			&H_{k} = (-t \cos k + \mu/2) \sigma_z + \Delta \sin k \sigma_y,\\
			&\epsilon_{k} = \sqrt{(t \cos k - \mu/2)^2 + (\Delta \sin k)^2}.
		\end{aligned}
	\end{equation}
	We can diagonalize $H_{k}$ and obtain eigenvectors $\Psi_{k, \pm}$ corresponding to eigenvalues $\pm \epsilon_{k}$:
	\begin{equation}
		\begin{aligned}
			&\Psi_{k, +} = \begin{pmatrix}
				\sqrt{\frac{\epsilon_{k} + (-t \cos k + \mu/2)}{2\epsilon_{k}}} \\
				\frac{\ii \Delta \sin k}{\sqrt{2\epsilon_{k}(\epsilon_{k} + (-t \cos k + \mu/2))}}
			\end{pmatrix},\\
			&\Psi_{k, -} = \begin{pmatrix}
				\frac{\ii \Delta \sin k}{\sqrt{2\epsilon_{k}(\epsilon_{k} + (-t \cos k + \mu/2))}}\\
				\sqrt{\frac{\epsilon_{k} + (-t \cos k + \mu/2)}{2\epsilon_{k}}}
			\end{pmatrix}.
		\end{aligned}
	\end{equation}
	The Hamiltonian can always be decomposed into two terms:
	\begin{equation}
		\begin{aligned}
			H_{k} &= \epsilon_{k}\Psi_{k, +}\Psi_{k, +}^{\dagger} - \epsilon_{k}\Psi_{k, -}\Psi_{k, -}^{\dagger}\\
			& = H_{k}^{(+)} + H_{k}^{(-)},
		\end{aligned}
	\end{equation}
	where we note that
	\begin{equation}
		\begin{aligned}
			H_{\bm{k}}^{(+)} &= \frac{\epsilon_{\bm{k}}}{2}\mathds{1}_2 + \frac{H_{\bm{k}}}{2},\\
			H_{\bm{k}}^{(-)} &= -\frac{\epsilon_{\bm{k}}}{2}\mathds{1}_2 + \frac{H_{\bm{k}}}{2}.
		\end{aligned}
	\end{equation}
	For a general parameter choice, $H_{\bm{k}}^{(+)}$ and $H_{\bm{k}}^{(-)}$ are not Laurent polynomials of $e^{\ii k}$.
	Therefore, the Hamiltonian is not frustration-free.
	
	However, as discussed in Sec.~\ref{sec:kitaev}, the Hamiltonian becomes frustration-free for a specific parameter choice: $(t, \Delta, \mu) = (t_0/2, -t_0 \sin 2\theta/2, t_0 \cos 2\theta)$ with $t_0 > 0, \theta \in \mathbb{R}$.
	For this choice, we obtain
	\begin{align}
		\epsilon_{k} = t_0 (1 - \cos 2 \theta \cos k)/2.
	\end{align}
	Consequently, $H_{\bm{k}}^{(\pm)}$ become
	\begin{align}
		H_{\bm{k}}^{(\pm)} = \pm [t_0 (1 - \cos 2 \theta \cos k)/2] \mathds{1}_2 + \frac{H_{\bm{k}}}{2},
	\end{align}
	which are indeed Laurent polynomials of $e^{\ii k}$.
	Furthermore, with an appropriate gauge choice, we find
	\begin{align}
		\Psi_{k, -} \propto \begin{pmatrix}
			\cos \theta (e^{\ii k} + 1) \\ \sin \theta (e^{\ii k} - 1)
		\end{pmatrix}.
	\end{align}
	Note that the unnormalized version of $\Psi_{k, -}$ is a Laurent polynomial of $e^{\ii k}$, implying that the corresponding Wannier-type function is compactly supported.

	\subsubsection{Chern insulator}
	Next, we consider the Chern insulator, whose Hamiltonian in momentum space is given by
	\begin{equation}
		\begin{aligned}
			&\hat{H} = \sum_{\bm{k}} \begin{pmatrix}
				\hat{c}_{\bm{k}, 1}^\dagger & \hat{c}_{\bm{k},2 }^\dagger
			\end{pmatrix} H_{\bm{k}} \begin{pmatrix}
				\hat{c}_{\bm{k}, 1} \\ \hat{c}_{\bm{k},2}
			\end{pmatrix},\\
			&H_{\bm{k}} = t (\sin k_x \sigma_x + \sin k_y \sigma_y) + M_{\bm{k}}\sigma_z\ \ (t \in \mathbb{R}),\\
			&M_{\bm{k}} = m+\sum_{i=1}^{2}\cos k_i\ \ (m \in \mathbb{R}).
		\end{aligned}
	\end{equation}
	The Chern number $\mathrm{Ch}_1$ is computed as
	\begin{align}
		\mathrm{Ch}_1 = \begin{cases}
			0 & |m| > 2, \\
			-\mathrm{sgn}\ m & 0 < |m| < 2. \\
		\end{cases}
	\end{align}
	It is well known that any set of bands with a nonzero Chern number does not admit compactly supported Wannier functions~\cite{PhysRevLett.98.046402}.
	In the following, we explicitly show that the Hamiltonian is not frustration-free.
	
	The energy dispersion $\epsilon_{\bm{k}}$ of $H_{\bm{k}}$ is
	\begin{align}
		\epsilon_{\bm{k}} = \sqrt{M_{\bm{k}}^2 + t^2(\sin^2  k_x + \sin^2 k_y)}.
	\end{align}
	The projections onto the occupied and unoccupied bands are given by
	\begin{align}
		P_{\bk, \pm} = \frac{1}{2} \mathds{1}_2 \pm \frac{H_{\bk}}{2 \epsilon_{\bm{k}}}.
	\end{align}
	Thus, $H_{\bk}$ can be decomposed into two terms:
	\begin{equation}
		\begin{aligned}
			&H_{\bk} = \epsilon_{\bm{k}} P_{\bk, +} - \epsilon_{\bm{k}} P_{\bk, -} = H_{\bk}^{(+)} + H_{\bk}^{(-)},\\
			&H_{\bk}^{(+)} = \frac{\epsilon_{\bm{k}}}{2}\mathds{1}_2 + \frac{H_{\bm{k}}}{2},\\
			&H_{\bk}^{(-)} = -\frac{\epsilon_{\bm{k}}}{2}\mathds{1}_2 + \frac{H_{\bm{k}}}{2}.
		\end{aligned}
	\end{equation}
	Although $H_{\bm{k}}$ is a Laurent polynomial, $\epsilon_{\bm{k}}$ is not for any parameter set $(m, t)$.
	Consequently, $H_{\bm{k}}^{(+)}$ and $H_{\bm{k}}^{(-)}$ are not Laurent polynomials.
	Therefore, the Hamiltonian is not frustration-free.  However, we note that if one relaxes the strict locality condition to allow for exponentially decaying hoppings, it is possible to construct a frustration-free parent Hamiltonian for the Chern insulator, as demonstrated in Ref.~\cite{Sengoku2025}.
	
	\subsubsection{Three-dimensional class AIII}
	As a final example, we consider a three-dimensional Hamiltonian in class AIII:
	\begin{equation}
		\begin{aligned}
			H_{\bm{k}} &=  t\left(\sin k_x \tau_x \sigma_x  + \sin k_y \tau_x \sigma_y+ \sin k_z \tau_x \sigma_z \right) + M_{\bm{k}}\tau_y\sigma_0,\\
			M_{\bm{k}} &= m+\sum_{i=1}^{3}\cos k_i,\\
			U_{\Gamma} &= \tau_z \sigma_0,
		\end{aligned}
	\end{equation}
	where $U_{\Gamma} H_{\bm{k}} = - H_{\bm{k}} U_{\Gamma}$.
	The three-dimensional winding number $w_{3\text{d}}$ of $H_{\bm{k}}$ is given by~\cite{shiozaki2024discreteformulationthreedimensionalwinding}
	\begin{align}
		w_{3\text{d}} = \begin{cases}
			0 & |m| > 3, \\
			-2\ \mathrm{sgn}\ t & |m| < 1, \\
			\mathrm{sgn}\ t & 1 < |m| < 3. \\
		\end{cases}
	\end{align}
	Due to accidental degeneracy, the magnitudes of all eigenvalues of $H_{\bm{k}}$ are identical and are given by
	\begin{equation}
		\epsilon_{\bm{k}} = \sqrt{M_{\bm{k}}^2 + t^2(\sin^2  k_x + \sin^2 k_y + \sin^2 k_z)}.
	\end{equation}
	The corresponding eigenvectors are
	\begin{equation}
		\begin{aligned}
			&\Psi_{\bm{k}, \pm}^{(1)} = \frac{1}{\sqrt{2} \epsilon_{\bm{k}}}\begin{pmatrix}
				t \sin k_z - \ii M_{\bm{k}} \\
				t (\sin k_x + \ii \sin k_y) \\
				\pm \epsilon_{\bm{k}} \\
				0
			\end{pmatrix}, \\
			&\Psi_{\bm{k}, \pm}^{(2)} = \frac{1}{\sqrt{2} \epsilon_{\bm{k}}}\begin{pmatrix}
				t(\sin k_x - \ii \sin k_y) \\
				-t \sin k_z - \ii M_{\bm{k}} \\
				0 \\
				\pm \epsilon_{\bm{k}}
			\end{pmatrix},
		\end{aligned}
	\end{equation}
	where $\pm$ corresponds to positive and negative eigenvalues.
	The eigenvectors are non-singular at any momentum for $|m| \neq 1, 3$; thus, they are analytic functions of $\bk$ in the gapped regime $|m| \neq 1, 3$.
	Therefore, the Wannier functions are exponentially localized.
	
	The projections onto the occupied and unoccupied bands are computed as
	\begin{align}
		P_{\bk, \pm} = \left( \Psi_{\bm{k}, \pm}^{(1)} \Psi_{\bm{k}, \pm}^{(2)} \right) \left( \Psi_{\bm{k}, \pm}^{(1)} \Psi_{\bm{k}, \pm}^{(2)} \right)^{\dagger} =\frac{1}{2} \mathds{1}_4 \pm \frac{H_{\bk}}{2 \epsilon_{\bm{k}}}.
	\end{align}
	Again, we decompose the Hamiltonian into
	\begin{equation}
		\begin{aligned}
			&H_{\bm{k}} = \epsilon_{\bm{k}} P_{\bk, +} - \epsilon_{\bm{k}} P_{\bk, -} = H_{\bk}^{(+)} + H_{\bk}^{(-)},\\
			&H_{\bk}^{(+)} = \frac{\epsilon_{\bm{k}}}{2}\mathds{1}_4 + \frac{H_{\bm{k}}}{2},\\
			&H_{\bk}^{(-)} = -\frac{\epsilon_{\bm{k}}}{2}\mathds{1}_4 + \frac{H_{\bm{k}}}{2}.
		\end{aligned}
	\end{equation}
	While $H_{\bm{k}}$ is a Laurent polynomial, $\epsilon_{\bm{k}}$ is not for any parameter $(m, t)$.
	Therefore, we conclude that $H_{\bm{k}}^{(+)}$ and $H_{\bm{k}}^{(-)}$ are not Laurent polynomials; i.e., the Hamiltonian is not frustration-free.

\subsection{Fragile Topology}
\label{sec:fragile}
Here we show that a fragile topology~\cite{PhysRevLett.121.126402} can be realized in frustration-free free-fermions.
Our example is a five-band model designed for the moir\'e superlattice in twisted bilayer graphene~\cite{PhysRevB.99.195455}. 
The lattice vectors are defined as $\bm{a}_1 = ( \frac{\sqrt{3}}{2}, -\frac{1}{2} )$ and $\bm{a}_2 = ( 0, 1 )$.
The triangular site ($\tau$) is located at $\bm{r} = \bm{0}$ and hosts three orbitals $p_z, p_+, p_-$ (where $p_\pm = p_x \pm i p_y$).
The honeycomb sites ($\eta$) are located at $\bm{r}_{A} = \frac{1}{3}\bm{a}_1+\frac{2}{3}\bm{a}_2$ and $\bm{r}_{B} = \frac{2}{3}\bm{a}_1+\frac{1}{3}\bm{a}_2$, and host an $s$-orbital on each sublattice.

\subsubsection{Symmetries}
The spatial symmetries are generated by $C_3$ rotation about the $z$-axis, $M_y$ mirror ($y \to -y$)~\footnote{As explained in Ref.~\cite{PhysRevB.99.195455}, $M_y$ is actually the two-fold rotation about the $x$ axis.}, and $C_2$ rotation about the $z$-axis combined with time-reversal $\mathcal{T}$.
The coordinate transformations acting on $\bm{r}=(x,y,z)$ are given by:
\begin{align}
R(C_3)&= \begin{pmatrix}
-\frac{1}{2} & -\frac{\sqrt{3}}{2} & 0 \\
\frac{\sqrt{3}}{2} & -\frac{1}{2} & 0 \\
0 & 0 & 1
\end{pmatrix},\\
R(M_y)&= \begin{pmatrix}
1 & 0 & 0 \\
0 & -1 & 0 \\
0 & 0 & -1
\end{pmatrix},\\
R(C_2\mathcal{T})&=\begin{pmatrix}
-1 & 0 & 0 \\
0 & -1 & 0 \\
0 & 0 & 1
\end{pmatrix}.
\end{align}

Due to the non-primitive positions of the honeycomb sites, the unitary matrices $U_{\bm{k}}(g)$ depend on $\bm{k}$. In the basis of $\Psi_{\bm{k}}^\dagger = (\hat{\tau}_{p_z}^\dagger, \hat{\tau}_{p_+}^\dagger, \hat{\tau}_{p_-}^\dagger, \hat{\eta}_{A}^\dagger, \hat{\eta}_{B}^\dagger)$, 
\begin{align}
    U_{\bm{k}}(C_3) &= 
    \begin{pmatrix} 
        1 & 0 & 0 & 0 & 0 \\
        0 & e^{i\frac{2\pi}{3}} & 0 & 0 & 0 \\
        0 & 0 & e^{i\frac{4\pi}{3}} & 0 & 0 \\
        0 & 0 & 0 & e^{-i\bm{k}\cdot\bm{a}_2} & 0 \\
        0 & 0 & 0 & 0 & e^{-i\bm{k}\cdot(\bm{a}_1+\bm{a}_2)}
    \end{pmatrix},\\
    U_{\bm{k}}(M_y) &= \begin{pmatrix} 
        -1 & 0 & 0 & 0 & 0 \\
        0 & 0 & 1 & 0 & 0 \\
        0 & 1 & 0 & 0 & 0 \\
        0 & 0 & 0 & e^{-i\bm{k}\cdot\bm{a}_2} & 0 \\
        0 & 0 & 0 & 0 & 1 
    \end{pmatrix},\\
    U_{\bm{k}}(C_2\mathcal{T}) &= \begin{pmatrix} 
        1 & 0 & 0 & 0 & 0 \\
        0 & 0 & 1 & 0 & 0 \\
        0 & 1 & 0 & 0 & 0 \\
        0 & 0 & 0 & 0 & e^{i\bm{k}\cdot(\bm{a}_1+\bm{a}_2)} \\
        0 & 0 & 0 & e^{i\bm{k}\cdot(\bm{a}_1+\bm{a}_2)}  & 0 
    \end{pmatrix}.
\end{align}
A symmetric Hamiltonian satisfies
\begin{align}
&U_{\bm{k}}(C_3) H_{\bm{k}} U_{\bm{k}}(C_3)^\dagger =H_{R(C_3)\bm{k}},\\
&U_{\bm{k}}(M_y) H_{\bm{k}} U_{\bm{k}}(M_y)^\dagger =H_{R(M_y)\bm{k}},\\
&U_{\bm{k}}(C_2\mathcal{T}) H_{\bm{k}}{}^* U_{\bm{k}}(C_2\mathcal{T})^\dagger =H_{\bm{k}}.
\end{align}
The five bands in total have symmetry representations summarized in the first row of Table~\ref{tab:fragile}.

\begin{table}[t]
\centering
\caption{Symmetry eigenvalues at high-symmetry points $\Gamma$, $M$, and $K$. We define $\omega \coloneqq e^{i2\pi/3}$.
}
\label{tab:fragile}
\renewcommand{\arraystretch}{1.2}
\begin{tabular}{lcccc}
\toprule
\multirow{2}{*}{Set of bands} & \multicolumn{2}{c}{$\Gamma=(0,0)$} & $M=(\frac{2\pi}{\sqrt{3}},0)$ & $K=(\frac{2\pi}{\sqrt{3}},\frac{2\pi}{3})$ \\
\cmidrule(lr){2-3} \cmidrule(lr){4-4} \cmidrule(lr){5-5}
& $C_3$ & $M_y$ & $M_y$ & $C_3$ \\ \cmidrule(lr){1-5}
\multirow{2}{*}{5 bands total} & $1,1,1,$ & $1,1,1,$ & $1,1,1,$ & $1,\omega,\omega,$ \\ 
~ & $\omega, \omega^*$ & $-1,-1$ & $-1,-1$ & $\omega^*,\omega^*$
\\ \cmidrule(lr){1-5}
Conduction  & $1, \omega, \omega^*$ & $1, 1, -1$ & $1, 1, -1$ & $1, \omega, \omega^*$ \\\cmidrule(lr){1-5}
Valence & $1,1$ & $1, -1$ & $1, -1$ & $\omega, \omega^*$ \\\cmidrule(lr){1-5}
Triangular $p_\pm$ & $\omega,\omega^*$ & $1, -1$ & $1, -1$ & $\omega, \omega^*$ \\
\bottomrule
\end{tabular}
\end{table}

\subsubsection{Conduction bands}
The Hamiltonian for the positive-energy bands is given by
\begin{align}
H_{\bm{k}}^{(+)} \coloneqq t_0 \rho_{\bm{k}} \rho_{\bm{k}}^\dagger=t_0^{(+)} \sum_{\alpha=1}^3\rho_{\bm{k}}^{(\alpha)}\rho_{\bm{k}}^{(\alpha)}{}^\dagger,
\end{align}
where
\begin{align}
&\rho_{\bm{k}}^{(1)} = \begin{pmatrix} i\tilde{a}(e^{-i\bm{k}\cdot\bm{a}_2} - 1) \\ \tilde{b}e^{-i\bm{k}\cdot\bm{a}_2} + \tilde{c} \\ \tilde{c}e^{-i\bm{k}\cdot\bm{a}_2} + \tilde{b} \\ \tilde{d}^* \\ \tilde{d}e^{i\bm{k}\cdot\bm{a}_1} \end{pmatrix}, \\
&\rho_{\bm{k}}^{(2)}=U_{\bm{k}'}(C_3)\rho_{\bm{k}'}^{(1)}\Big|_{\bm{k}'=R(C_3)^{-1}\bm{k}},\\
&\rho_{\bm{k}}^{(3)}=U_{\bm{k}'}(C_3)\rho_{\bm{k}'}^{(2)}\Big|_{\bm{k}'=R(C_3)^{-1}\bm{k}}
\end{align}
corresponds to quasi-orbitals for $s$-orbitals centered at Kagome sites 
$\bm{r}_{1}= \frac{1}{2}\bm{a}_2$, $\bm{r}_{2}=-\frac{1}{2}\bm{a}_1-\frac{1}{2}\bm{a}_2$, $\bm{r}_{3}=\frac{1}{2}\bm{a}_1$ as illustrated in Fig.~\ref{figfragile}(a). Note that our choice of $\rho_{\bm{k}}^{(1,2,3)}$ differs from that in Ref.~\cite{PhysRevB.99.195455} by a phase factor. Here $t_0, \tilde{a},\tilde{b},\tilde{c}$ are real and $\tilde{d}$ is a complex parameter.  These are aggregated into $\rho_{\bm{k}} = \begin{pmatrix}\rho_{\bm{k}}^{(1)}& \rho_{\bm{k}}^{(2)}& \rho_{\bm{k}}^{(3)}\end{pmatrix}$. 

If we set $t_0=\tilde{a}=\tilde{b}=\tilde{c}=\tilde{d}=1$, for example, the conduction bands have energy $4\leq\epsilon_{\bm{k}}\leq 12$ and have symmetry representations summarized in the second row of Table~\ref{tab:fragile}. 

\subsubsection{Valence bands}
The symmetry representations of the valence bands are uniquely determined by subtracting the representations of the conduction bands from those of the total set of atomic orbitals. The results are summarized in the third row of Table~\ref{tab:fragile}. Crucially, there are no atomic orbitals that have the same symmetry representations as these valence bands. This implies the fragile topology in the valence bands.

\begin{figure}[t]
\begin{center}
\includegraphics[width=0.5\textwidth]{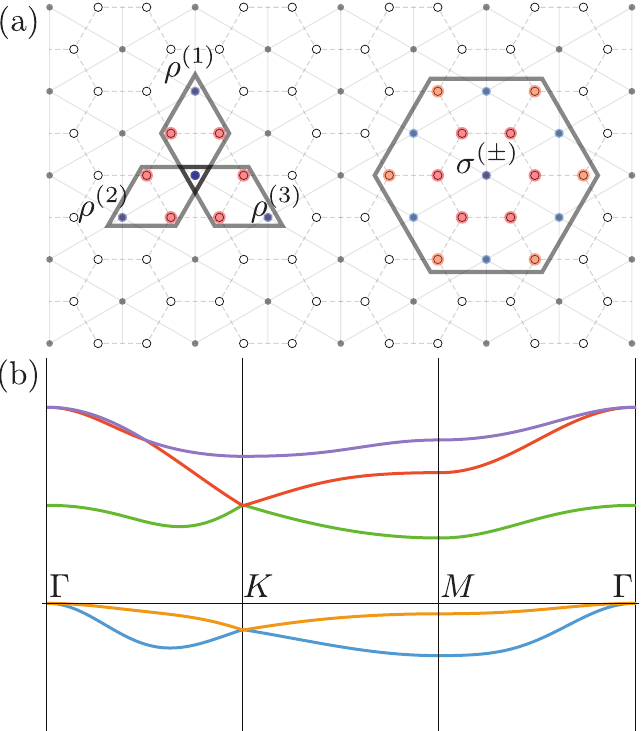}
\caption{ 
(a) Quasi-orbitals for conduction bands ($\rho^{(1,2,3)}\coloneqq\sum_{\bm{k}}\rho_{\bm{k}}^{(1,2,3)}e^{i\bm{k}\cdot\bm{r}}$) and for valence bands ($\sigma^{(\pm)}\coloneqq\sum_{\bm{k}}\sigma_{\bm{k}}^{(\pm)}e^{i\bm{k}\cdot\bm{r}}$).
(b) The band dispersion of $H_{\bm{k}}=H_{\bm{k}}^{(+)}+H_{\bm{k}}^{(-)}$.
\label{figfragile}
 }
\end{center}
\end{figure}

Nonetheless, we find it is still possible to define an \emph{incomplete} set of Wannier orbitals and give a dispersion to the valence bands within frustration-free free-fermions. To construct the Hamiltonian for negative-energy bands,  we write
\begin{align}
\rho_{\bm{k}}^\dagger = \begin{pmatrix} \bm{c}_1 & \bm{c}_2 & \bm{c}_3 & \bm{c}_4 & \bm{c}_5 \end{pmatrix}
\end{align}
and construct the orthogonal states using a generalized cross-product method (analogous to Cramer's rule). We find
\begin{align}
&\sigma_{\bm{k}}^{(+)} 
\coloneqq
ie^{i\bm{k}\cdot(\bm{a}_1+\bm{a}_2)}
\begin{pmatrix}
\det(\bm{c}_2, \bm{c}_4, \bm{c}_5) \\ 
-\det(\bm{c}_1, \bm{c}_4, \bm{c}_5) \\
0\\ 
\det(\bm{c}_1, \bm{c}_2, \bm{c}_5) \\ 
-\det(\bm{c}_1, \bm{c}_2, \bm{c}_4) 
\end{pmatrix},\\
&\sigma_{\bm{k}}^{(-)} 
\coloneqq U_{\bm{k}}(C_2\mathcal{T})\Big(\sigma_{\bm{k}}^{(+)}\Big)^*
\end{align}
correspond to $p_+, p_-$-orbitals centered at the triangular sites $\bm{r}=\bm{0}$ (Fig.~\ref{figfragile}(a)). Note that $\sigma_{\bm{k}}^{(\pm)}$ are orthogonal to $\rho_{\bm{k}}^{(1,2,3)}$.  The Hamiltonian for the negative bands is then given by
\begin{align}
&H_{\bm{k}}^{(-)} \coloneqq -t_0' \sigma_{\bm{k}}\sigma_{\bm{k}}^\dagger,\quad \sigma_{\bm{k}}\coloneqq \begin{pmatrix} \sigma_{\bm{k}}^{(+)}& \sigma_{\bm{k}}^{(-)}\end{pmatrix}.
\end{align}
Since the rotation eigenvalues of the valence bands do not agree with $p_\pm$ orbitals centered at the triangular site (compare the third and fourth rows in Table~\ref{tab:fragile}), the induced dispersion vanishes at $\Gamma$ (Fig.~\ref{figfragile}(b)).

The existence of the compact localized states $\sigma^{(\pm)}$ does not contradict the fragile nature of the topology. We find that the Wannier states corresponding to $\sigma_{\bm{k}}^{(\pm)}$ are linearly dependent, and the Hilbert space is completed by two extended states: $\hat{\tau}_{s,\bm{k}=\bm{0}}^\dagger|0\rangle$ and $(\hat{\eta}_{A,p_+,\bm{k}=\bm{0}}^\dagger-\hat{\eta}_{B,p_-,\bm{k}=\bm{0}}^\dagger)|0\rangle$.

\section{Gosset-Huang inequality and its implication}
\label{sec:GH}
In this section, we discuss the scaling of finite-size gaps in frustration-free systems assuming power-law behaviors of correlation functions.

\subsection{Statement}

Let $E_n$ ($n=1,2,\cdots$) be the many-body energy eigenvalues of $\hat{H}$ arranged in the increasing order:
\begin{align}
E_1\leq E_2\leq \cdots.
\end{align}
We assume $E_1=E_2=\cdots=E_{N_{\mathrm{deg}}}=0$ and $E_{N_{\mathrm{deg}}+1}>0$. Then the finite-size gap is defined as
\begin{align}
\epsilon=E_{N_{\mathrm{deg}}+1}-E_{N_{\mathrm{deg}}}.
\end{align}
In gapless systems, $\epsilon$ vanishes in the thermodynamic limit. Namely, if $L$ is the linear size of the system, $\lim_{L\to\infty}\epsilon=0$ in gapless systems and $\lim_{L\to\infty}\epsilon>0$ in gapped systems~\cite{arXiv:2406.06414}.

We consider local operators $\hat{\mathcal{O}}_{\bm{x}}$ and $\hat{\mathcal{O}}_{\bm{y}}'$ acting nontrivially around sites $\bm{x}$ and $\bm{y}$, respectively. For any ground state $|\Phi\rangle$ of a frustration-free Hamiltonian, the correlation function between $\hat{\mathcal{O}}_{\bm{x}}$ and $\hat{\mathcal{O}}_{\bm{y}}'$ satisfies the following Gosset--Huang inequality~\cite{gossetCorrelationLengthGap2016}: 
\begin{align} &
 \frac{|\langle\Phi|\hat{\mathcal{O}}_{\bm{x}}(\hat{\mathbbm{1}}-\hat G)\hat{\mathcal{O}}'_{\bm{y}}|\Phi\rangle|}
 	{\|\hat{\mathcal{O}}^\dagger_{\bm{x}}|\Phi\rangle\| \|\hat{\mathcal{O}}'_{\bm{y}}|\Phi\rangle\|} \nonumber\\
	&
  \leq 2e^2 \exp(-\frac{D(\hat{\mathcal{O}}_{\bm{x}},\hat{\mathcal{O}}_{\bm{y}}')-1}{c-1}\sqrt{\frac{\epsilon}{g^2+\epsilon}}),
 \end{align}
where $\hat G$ is the projector onto the ground space, $D(\hat{\mathcal{O}}_{\bm{x}},\hat{\mathcal{O}}_{\bm{y}}')$ is the distance between the two operators, and $c$ and $g$ are integer constants determined by details of the Hamiltonian.
We include the accurate definitions of $D(\cdot, \cdot)$, $c$ and $g$ in Appendix~\ref{app:GH}.
The proof of this inequality is in Ref.~\cite{gossetCorrelationLengthGap2016} and Ref.~\cite{arXiv:2406.06415}.
Although this inequality was originally derived for bosonic systems, it can be extended to fermionic systems with bosonic local Hamiltonians.  Importantly, the local operators $\hat{\mathcal{O}}_{\bm{x}}$ and $\hat{\mathcal{O}}_{\bm{y}}'$ do not need to be bosonic.

Suppose that operators $\hat{\mathcal{O}}_{\bm{x}}$ and $\hat{\mathcal{O}}_{\bm{y}}'$ with the distance $D(\hat{\mathcal{O}}_{\bm{x}}, \hat{\mathcal{O}}_{\bm{y}}') \sim L$ shows power-law decay
\begin{align}
\frac{|\langle\Phi|\hat{\mathcal{O}}_{\bm{x}}(\hat{\mathbbm{1}}-\hat G)\hat{\mathcal{O}}'_{\bm{y}}|\Phi\rangle|}
{\|\hat{\mathcal{O}}_{\bm{x}}|\Phi\rangle\| \|\hat{\mathcal{O}}'_{\bm{y}}|\Phi\rangle\|} \geq CL^{-\Delta} \label{assumption for correlation functions}
\end{align}
for positive constants $C>0$ and $\Delta>0$. 
Then the Gosset--Huang inequality implies that the finite-size gap satisfies~\cite{arXiv:2406.06415}
\begin{align}
\epsilon \leq C'\frac{(\log L)^2}{L^2},\label{GHbound}
\end{align}
where $C'$ is a positive constant. 

A finite energy gap in the thermodynamic limit guarantees exponential decay of correlation functions in the ground state, for both bosonic and fermionic systems \cite{hastingsSpectralGapExponential2006}. Therefore the correlation function in Eq.~\eqref{assumption for correlation functions} in the ground state implies that the Hamiltonian is gapless.
In Appendix \ref{app:power-law}, we provide two free-fermion examples of such power-law decaying correlation functions, and show that the existence of a singular band touching point between the valence and conduction bands imply power-law decaying ground-state correlation functions.

\subsection{Examples}
\subsubsection{Tight-binding model on checkerboard lattice}
Let us first check the validity of the bound~\eqref{GHbound} using the tight-binding model. As an example, we take the tight-binding model on the checkerboard lattice discussed in Sec.~\ref{sec:Checkerboard lattice}. Let $|\Phi\rangle$ be the ground state in which all states in the lower band, except for the one at $\bm{k}=\bm{0}$, are occupied. The state shows the power-law correlation
\begin{align}
	\langle\Phi|\hat a_{\bm{0}}(\hat{\mathbbm{1}}-\hat G)\hat a^\dagger_{\bm{R}}|\Phi\rangle\approx \frac1{2\pi}\frac{R_y^{\,2} - R_x^{\,2}}{|\bm{R}|^4}(-1)^{R_x+R_y}
\end{align}
asymptotically for a large $|\bm{R}|$.
See Appendix~\ref{app:power-law} for the derivation. Combined with $\|\hat a^\dagger_{\bm{0}}|\Phi\rangle\| \approx 1/\sqrt{2}$, we find the desired power-law decay. Then the bound~\eqref{GHbound} implies a low-energy excitation, which is consistent with the quadratic band touching discussed in Sec.~\ref{TRinv}.

\subsubsection{$\eta$-pairing state}
The bound~\eqref{GHbound} applies not only to free fermions but also to interacting fermions with or without the translation invariance. For example, the $\eta$-pairing state defined on a $d$-dimensional hypercubic lattice exhibits an off-diagonal long-rage order~\cite{RevModPhys.34.694}, which may be written as 
\begin{align}
\frac{\langle\Phi_N|\hat{\eta}_x^\dagger(\hat{\mathbbm{1}}-\hat{G})\hat{\eta}_y|\Phi_N\rangle}{\|\hat{\eta}_x|\Phi_N\rangle\|\|\hat{\eta}_y|\Phi_N\rangle\|}=-\frac{N-1}{V(V-1)}.
\end{align}
Here $V=L^d$ is the system size and $N$ is the number of $\eta$-pairs. For some parameter range, these $\eta$-pairing states become ground states of a generalized Hubbard model and the model is frustration-free~\cite{PhysRevLett.68.2960,PhysRevLett.75.4298, PhysRevB.105.024520}. The power-law correlation function then indicates the presence of gapless excitations bounded by \eqref{GHbound}.

\section{Conclusion}
In this work, we have studied frustration-free fermionic systems and established fundamental properties that characterize their low-energy excitations. For free fermionic systems, we have derived a necessary and sufficient condition for frustration-freeness.
In particular, we have demonstrated that, in the presence of translation symmetry, frustration-free free fermionic systems exhibit quadratic dispersions for low-energy excitations.
Furthermore, we have generalized our findings to interacting and non-translation-invariant fermionic systems by extending the Gosset-Huang inequality.
Specifically, we have shown that, if the ground-state correlation function decays as a power law, the finite-size scaling of gapless excitations is $O((\log L)^2/L^2)$ even in the presence of interactions and/or the absence of translation symmetry. This result implies, among other things, that the effective field theory of gapless frustration-free fermionic systems cannot be of Dirac type; in particular, it must break Lorentz invariance.


\begin{acknowledgments}
We thank Hosho Katsura, Ken Shiozaki, Hal Tasaki, Carolyn Zhang, and Ruben Verresen for useful discussions.
H.W., S.O., and H.C.P.~thank the Yukawa Institute for Theoretical Physics, Kyoto University, where we started the project during the YITP workshop YITP-T-24-03 on ``Recent Developments and Challenges in Topological Phases''.
The work of H.W. is supported by JSPS KAKENHI Grant No.~JP24K00541.
S.O.~was supported by RIKEN Special Postdoctoral Researchers Program.
H.C.P.~acknowledges support from the National Key R\&D Program of China (Grant No. 2021YFA1401500) and the Hong Kong Research Grants Council (C7037-22GF).
\end{acknowledgments}

\appendix

\section{Analytic property of band dispersion}
\label{app:analytic}
In this appendix, we discuss the analytic property of band dispersion $\omega_{\bm{k}n}$.

In one dimension, it is known that eigenvalues of $H(k)$ can be labeled in such a way that $\omega_{kn}$ ($i=1,2,\cdots,N_{\mathrm{band}}$) are all power series of $k$ convergent for small $|k|$, as long as each component of $H(k)$ is a power series of $k$~\cite{rellich1969perturbation}.
To see this, recall that eigenvalues of $H(k)$ are roots of the $N_{\mathrm{band}}$-th order polynomial
\begin{align}
P(\omega)=\det(H(k)-\omega I_{N_{\mathrm{band}}})
\end{align}
Then the Newton-Puiseux theorem suggests that every root of this equation can be expressed in the form of the Puiseux series (a generalization of the Laurant series)~\cite{Basu},
\begin{align}
\omega=\sum_{m\geq m_0}c_mk^{m/q},
\end{align}
where $m_0$ is an integer (can be negative), $q$ is a positive integer, $m\in\mathbb{Z}$, and $c_m\in\mathbb{C}$. For a given polynomial, a step-by-step procedure of finding roots in the Puiseux series form using the Newton polygon is known.

Since $H(k)$ is Hermitian and its dependence on $k$ is analytic, $\omega$ must be real and finite for any $k$. This is possible only when all powers appearing in the Puiseux series are nonnegative integers.

In higher dimension, this argument applies to any one-dimensional curve $\bm{k}(t)$ in the Brillouin zone.

\section{Alternative discussions for U(1) breaking systems}
\label{app:BdG}
In this appendix, we clarify how the frustration-free free fermion condition interacts with the BdG quasi-particle spectrum.

\subsection{BdG quasi-particle energy}
We start with a verbose recap of the ``local diagonalization'' discussion in the main text.

Consider Nambu spinor 
\begin{equation}
\hat {\bm{\Upsilon}} = (\hat c_1, \hat c_2,\dots, \hat c_N, \hat c_1^\dagger, \hat c_2^\dagger, \dots, \hat c_N^\dagger)^\top.
\end{equation}
Component-wise, we have
\begin{equation}
    \hat \Upsilon_i = \left\{
    \begin{array}{ll}
         \hat c_i &  \text{if }i\leq N,\\
         \hat c_{i-N+1}^\dagger & \text{otherwise.}
    \end{array}
    \right.
\end{equation}
Notice the anticommutation relation 
\begin{equation}
    \{ \hat \Upsilon_i, \hat \Upsilon_j^\dagger \} = \delta_{ij}.
\end{equation}
However, one cannot interpret $\hat \Upsilon$ as $2N$ independent canonical fermion modes because 
\begin{equation}
    \{ \hat \Upsilon_i, \hat \Upsilon_j \} = (U_{\mathcal P})_{ij},
\end{equation}
where 
\begin{equation}
    U_{\mathcal P} = 
    \left(
    \begin{array}{cc}
         0 & \mathbbm{1} \\
         \mathbbm{1} & 0
    \end{array}
    \right)
\end{equation}
is the unitary part of the anti-unitary particle-hole symmetry transforming 
\begin{equation}
\hat {\bm{\Upsilon}} \mapsto \hat {\bm{\Upsilon}}^* = U_{\mathcal P}^\dagger \hat {\bm{\Upsilon}},
\end{equation}
here, the $*$ on $\hat {\bm{\Upsilon}}$ means that we replace each component in the column vector $\hat {\bm{\Upsilon}}$ by its Hermitian conjugate.

We now consider a free-fermion Hamiltonian $\hat H = \hat {\bm{\Upsilon}}^\dagger H_{\rm BdG} \hat {\bm{\Upsilon}}$, where the BdG Hamiltonian $H_{\rm BdG}$ satisfies the particle-hole symmetry
\begin{equation}
    U_{\mathcal P} H^*_{\rm BdG} U_{\mathcal P}^\dagger = - H_{\rm BdG}.
\end{equation}
Let $\bm{\psi}$ be an eigenvector of $h_{\rm BdG}$ with eigenvalue $E>0$. By particle-hole symmetry, we have
\begin{equation}
    H_{\rm BdG} (U_{\mathcal P} \bm{\psi}^*)
    = - U_{\mathcal P} H_{\rm BdG}^* U_{\mathcal P}^\dagger (U_{\mathcal P} \bm{\psi}^*) = - U_{\mathcal P} ( E \bm{\psi})^*,
\end{equation}
which implies $U_{\mathcal P}\bm{\psi}^*$ is another eigenvector of $H_{\rm BdG}$ with eigenvalue $-E$. As eigenvectors with distinct eigenvalues $\pm E$, we also have
\begin{equation}\begin{split}\label{eq:canonical_condition}
\bm{\psi}^\dagger U_{\mathcal {P}} \bm{\psi}^* = 0.
\end{split}\end{equation}

Now, we extend the discussion to all eigenvectors. Let $\psi$ be the $2N \times N^{(+)}$ matrix with columns being the eigenvectors associated with the $N^{(+)} \leq N$ positive eigenvalues. The eigenvectors can be orthonormalized such that $\psi^\dagger \psi = \mathbbm{1}_{N^{(+)}}$. We denote the $i$-th column of $\psi$ by $\bm{\psi}_i$. The BdG Hamiltonian can be diagonalized through 
\begin{equation}\begin{split}\label{eq:}
H_{\rm BdG}
\left(
\begin{array}{cc}
     \psi & U_{\mathcal{P}} \psi^*
\end{array}
\right)
= \left(
\begin{array}{cc}
     \psi & U_{\mathcal{P}} \psi^*
\end{array}
\right)
\left(
\begin{array}{cc}
     D_E &  \\
     & -D_E
\end{array}
\right),
\end{split}\end{equation}
where $D_E$ is a diagonal matrix of the $N^{(+)}$ positive eigenvalues. As usual, the ``matrix diagonalization'' helps diagonalize the many-body Hamiltonian by noticing
\begin{equation}
\label{eq:diagonalized_bdg}
    \hat H = \hat {\bm {\Upsilon}}^\dagger 
    \left(
\begin{array}{cc}
     \psi & U_{\mathcal{P}} \psi^*
\end{array}
\right)
\left(
\begin{array}{cc}
     D_E &  \\
     & -D_E
\end{array}
\right)
\left(
\begin{array}{c}
     \psi^\dagger \\
     \psi^\top U_{\mathcal {P}}^\dagger
\end{array}
\right)
\hat {\bm {\Upsilon}}.
\end{equation}
The diagonalizing unitary satisfies
\begin{align}
    \left(
\begin{array}{c}
     \psi^\dagger \\
     \psi^\top U_{\mathcal {P}}^\dagger
\end{array}
\right)
\left(
\begin{array}{cc}
     \psi & U_{\mathcal{P}} \psi^*
\end{array}
\right)
&= 
\left(
\begin{array}{cc}
\psi^\dagger \psi & \psi^\dagger U_{\mathcal{P}} \psi^*\\
\psi^\top U_{\mathcal {P}}^\dagger\psi & \psi^\top \psi^*
\end{array}
\right)\notag\\
&=\mathbbm{1}_{2N^{(+)}}.
\end{align}
Notice the off-diagonal block gives the generalization of Eq.~\eqref{eq:canonical_condition}.

Let us define, for $i=1,\dots, N^{(+)}$,
\begin{equation}
    \hat \psi_i \coloneqq \bm{\psi}_i^\dagger \hat{\bm{\Upsilon}} = \sum_{j=1}^{2N} (\psi^\dagger)_{ij} \hat \Upsilon_j,
\end{equation}
which has Hermitian conjugate
\begin{equation}
    \hat \psi_i^\dagger = \bm{\psi}_i^\top \hat{\bm{\Upsilon}}^* 
     = \bm{\psi}_i^\top U_{\mathcal P}^\dagger \hat{\bm{\Upsilon}} 
    = \sum_{j=1}^{2N} \psi_{ji} \hat \Upsilon_j^\dagger
\end{equation}
We can verify the anti-commutation relation
\begin{equation}\begin{split}\label{eq:}
\{ \hat \psi_i, \hat \psi_j^\dagger \} = &
    \sum_{l,m} (\psi^\dagger)_{il} \psi_{mj}\{ \hat \Upsilon_l, \hat \Upsilon_m^\dagger \}\\
    = & \sum_{l,m}(\psi^\dagger)_{il} \psi_{mj} \delta_{lm} 
    = \sum_{l} (\psi^\dagger)_{il} \psi_{lj}
    = \delta_{ij}.
\end{split}\end{equation}
In addition, we have 
\begin{equation}\begin{split}\label{eq:}
\{ \hat \psi_i, \hat \psi_j \} = &
    \sum_{l,m} (\psi^\dagger)_{il} (\psi^\dagger)_{jm}\{ \hat \Upsilon_l, \hat \Upsilon_m\}\\
    = & \sum_{l,m}(\psi^\dagger)_{il} \psi^*_{mj} (U_{\mathcal{P}})_{lm} 
    = (\psi^\dagger U_{\mathcal P} \psi^*)_{ij} =0,
\end{split}\end{equation}
through the matrix-form of Eq.~\eqref{eq:canonical_condition}.
As such, $\{ \hat \psi_i\}_{i=1}^{N^{(+)}}$ is a set of $N^{(+)}$ canonical fermions.

Combined, we conclude Eq.~\eqref{eq:diagonalized_bdg} can be rewritten as
\begin{equation}
    \hat H = \sum_{i=1}^N E_i (\hat \psi_i^\dagger \hat \psi_i - \hat \psi_i \hat \psi_i^\dagger)
    = \sum_{i=1}^N 2 E_i  \hat \psi_i^\dagger \hat \psi_i  + \text{constant}.
\end{equation}
As $\hat \psi_i$ are canonical, $\hat H$ is now a sum of commuting terms and the full many-body spectrum can be obtained. Furthermore, rewriting in terms of the Nambu spinor $\hat {\bm {\Upsilon}}$ one can always write
\begin{equation}\begin{split}\label{eq:non_bdg_form}
\hat H = \hat {\bm {\Upsilon}}^\dagger H^{(+)}\hat {\bm {\Upsilon}},
\end{split}\end{equation}
where
\begin{equation}
    H^{(+)} = \psi D_E \psi^\dagger
\end{equation}
is a positive semidefinite matrix of dimension $2N\times 2N$ and rank $N^{(+)} \leq N$.

Two comments are in order: first, when we restrict ourselves to a local term in the Hamiltonian, the above is equivalent to the main text with the replacement $2E_i \rightarrow \mu_i$. Second, the discussion applies equally well when we think of $H_{\rm BdG}$ as being the BdG Hamiltonian for the full system. In particular, when the system has translation symmetry one can first block-diagonalize in momentum blocks coupling only states at $\pm \bm{k}$. The discussion above holds equally well, and so the quasi-particle spectrum of $\hat H$ at $\bm{k}$ is given by the Fourier transform of the matrix $H^{(+)}(\bm{k})$. Importantly, $H^{(+)}$ as defined (through diagonalization of $H_{\rm BdG}$) is not a local matrix in general, and so the spectrum of $H^{(+)}(\bm{k})$ can be non-analytic, say featuring only the upper Dirac cone.

\subsection{Frustration-free condition}
The discussion above can be applied to a local term $\hat H_{\bm{R}}$ of the Hamiltonian. We can now define a local matrix $H^{(+)}_{\bm{R}}$ in the same way as above, such that
\begin{equation}
    \hat H_{\bm {R}} = \hat {\bm{\Upsilon}}^\dagger H^{(+)}_{\bm{R}}\hat {\bm{\Upsilon}}; \quad
    H^{(+)}_{\bm{R}} = \psi_{\bm{R}} D_{\mu \bm{R}}\psi_{\bm{R}}^\dagger.
\end{equation}
Here, $\psi_{\bm{R}}$ is a $2N \times A_{\bm {R}}$ matrix when there are in total $N$ complex fermion modes in the full system and $A_{\bm {R}}$ positive eigenvalues for the local BdG Hamiltonian. Each column of $\psi_{\bm{R}}$ is a corresponding eigenvector of the local BdG Hamiltonian padded suitably with zeroes outside of the region. $D_{\mu \bm{R}}$ is the diagonal matrix of the positive eigenvalues $\mu_{\bm{R}\alpha}> 0$. The full Hamiltonian can be written as 
\begin{equation}\begin{split}\label{eq:many-body_Hloc}
\hat H = \hat {\bm{\Upsilon}}^\dagger H_{\rm{loc}}^{(+)}\hat {\bm{\Upsilon}}; \quad H_{\rm{loc}}^{(+)} = \sum_{\bm{R}} H^{(+)}_{\bm{R}} 
= \psi D_{\mu} \psi^\dagger,
\end{split}\end{equation}
where $D_{\mu} = \oplus_{\bm{R}} D_{\mu \bm{R}}$ and $\psi$ is the corresponding matrix obtained by stacking  $\psi_{\bm{R}}$ horizontally.
Any local BdG Hamiltonian can be expressed in this way.

However, generally speaking, the eigenspectrum of $H_{\rm{loc}}^{(+)}$ so-defined does not coincide with the quasiparticle spectrum of $\hat H$. For instance, $H_{\rm{loc}}^{(+)}$ is a positive semidefinite matrix with only local terms. If the system has translation invariance, we can analyze its spectrum in the momentum space, and it follows from the Dirac trick that any touching of the eigenspectrum of $H_{\rm{loc}}^{(+)}$ with $0$ must be quadratic or softer in momentum. This holds even if the true quasiparticle spectrum of $\hat H$ has zero-energy Dirac cones. 

We now show that the frustration-free condition implies the diagonalization of $H_{\rm loc}^{(+)}$ reproduces the true quasiparticle spectrum of $\hat H$. For convenience, let
\begin{equation}\begin{split}\label{eq:}
F \coloneqq \psi \sqrt{D_{\mu}} \overset{\rm SVD}{=} \xi \sigma \zeta^\dagger.
\end{split}\end{equation}
Here, we keep only the $N^{(+)}$ non-zero singular values, where $N^{(+)} \leq \min(2N, \sum_{\bm{R}}A_{\bm{R}})$. 
The singular value decomposition (SVD) of $F$ also diagonalizes $H^{(+)}_{\rm loc} = \xi \sigma^2 \xi^\dagger$. 
For $i=1,2,\dots, N^{(+)}$, define 
\begin{equation}
\hat \xi_i \coloneqq \sum_{l=1}^{2N}(\xi^\dagger)_{il} \hat \Upsilon_l.    
\end{equation}
As $\xi$ is unitary, $\{\hat \Upsilon_l, \hat \Upsilon_m^\dagger\} = \delta_{lm}$ implies $\{\hat \xi_i, \hat \xi_j^\dagger\} = \delta_{ij}$. Furthermore, recall we also defined $\hat \psi_{\alpha} \coloneqq \sum_{l} (\psi^\dagger)_{\alpha l} \hat \Upsilon_l$. We may also express $\hat \xi_i$ as a superposition of $\hat \psi$'s using $F^\dagger = \zeta \sigma \xi^\dagger \Rightarrow \xi^\dagger = \sigma^{-1} \zeta^\dagger \sqrt{D_\mu}\psi^\dagger$, and so
\begin{equation}\begin{split}\label{eq:}
\hat \xi_{i} \coloneqq  \sum_{l=1}^{2N}\left(\frac{1}{\sigma} \zeta^\dagger \sqrt{D_\mu}\psi^\dagger \right)_{il} \hat \Upsilon_l  = \sum_{j} M_{i\alpha} \hat \psi_\alpha,
\end{split}\end{equation}
where $M\coloneqq \sigma^{-1} \zeta^\dagger \sqrt{D_\mu}$. This implies 
\begin{equation}
    \{ \hat \xi_i, \hat \xi_j\} = M_{i\alpha} M_{j\beta} \{ \hat \psi_\alpha, \hat \psi_\beta \}.
\end{equation}
At this step, we invoke the frustration-free condition~\eqref{eq:BdG_conditiontobeFFFF}, which gives $\{ \hat \psi_\alpha, \hat \psi_\beta \} = 0$, and therefore we conclude $\{\hat \xi_{i}~:~ i =1,\dots, N^{(+)}\}$ is a set of canonical fermions. We can therefore rewrite Eq.~\eqref{eq:many-body_Hloc} as
\begin{equation}
    \hat H = \sum_{i=1}^{N^{(+)}} \sigma_i^2 \hat n_i,
\end{equation}
where $\hat n_i \coloneqq \hat \xi_i^\dagger \hat \xi_i$ are mutually commuting and therefore $\sigma_i^2$ can be identified with the positive eigenvalues of the original BdG Hamiltonian.

The discussion above leaves us with a natural counting problem: how big can $N^{(+)}$ be? By the assertion that $\{ \hat \xi\}$ is a set of canonical fermions, we must have $N^{(+)} \leq N$, the total number of complex fermions in the system. This property may not be evident in the above in which $N^{(+)}$ is introduced as the number of non-zero singular values of the $2N \times  \sum_{\bm{R}}A_{\bm{R}}$ matrix $F$. Yet, we note that 
\begin{equation}\begin{split}\label{eq:}
0 = \{ \hat \xi_i, \hat \xi_j\} 
=& \sum_{l,m=1}^{2N}(\xi^\dagger)_{il}(\xi^\dagger)_{jm} \{ \hat \Upsilon_l, \hat \Upsilon_m\} \\
=& \sum_{l,m=1}^{2N}(\xi^\dagger)_{il}(\xi^\dagger)_{jm} (U_{\mathcal P})_{lm} \\
=& \left(\xi^\dagger U_{\mathcal P} \xi^*\right)_{ij}.
\end{split}\end{equation}
Therefore, each of the $N^{(+)}$ columns of the matrix $U_{\mathcal P} \xi^*$ is an independent, normalized null vector of $H^{(+)}_{\rm loc} = \xi \sigma^2 \xi^\dagger$. This implies $2 N^{(+)} \leq {\rm dim}(H^{(+)}_{\rm loc}) = 2N$, and so $N^{(+)} \leq N$, as is required by consistency of the discussion. Furthermore, if $N^{(+)} \neq N$ then the mismatch manifests as zero-energy modes in the quasiparticle spectrum of $\hat H$, say when 
Therefore, each of the $N^{(+)}$ columns of the matrix $U_{\mathcal P} \xi^*$ is an independent, normalized null vector of $H^{(+)}_{\rm loc} = \xi \sigma^2 \xi^\dagger$. This implies $2 N^{(+)} \leq {\rm dim}(H^{(+)}_{\rm loc}) = 2N$, and so $N^{(+)} \leq N$, as is required by consistency of the discussion. Furthermore, if $N^{(+)} \neq N$ then the mismatch manifests as zero modes of $\hat H$, say those corresponding to a zero-energy quasiparticle flat band in the BdG spectrum.

\section{Derivation of Eq.~\eqref{eq:projector_relation}}
\label{app:proof_lemma_projector}
In this appendix, we derive Eq.~\eqref{eq:projector_relation}, whose statement is as follows:
Let $\{|f_n \rangle \coloneqq \hat{f}^{\dagger}_{n}\ket{0}\}_{n=1}^{M}$ be a set of one-particle states that are orthonormal to each other.
We also define the space $\mathcal{W}_{f}$ and an orthogonal projector by $\mathcal{W}_{f} \coloneqq \mathrm{Span} \left\{|f_n \rangle \right\}_{n=1}^{M}$ and $P_{\mathcal{W}_f} \coloneqq \sum_{n=1}^{M} |f_n \rangle \langle f_n |$.
Suppose that a ground state $|\Phi\rangle$ is the Slater determinant of $f$-orbitals, i.e., $|\Phi\rangle = \hat{f}^{\dagger}_{1} \hat{f}^{\dagger}_{2} \cdots \hat{f}^{\dagger}_{M} \ket{0}$.
Then, the projector is nothing but the one-particle correlation function, i.e.,
\begin{equation*}
	\begin{aligned}
		[P_{\mathcal{W}_f}]_{ij} = \langle i| P_{\mathcal{W}_f} |j \rangle = \langle \Phi | \hat{c}^{\dagger}_{j} \hat{c}_{i} | \Phi \rangle.
	\end{aligned},
\end{equation*}
where $\{|i\rangle \coloneqq \hat{c}^{\dagger}_i | 0 \rangle\}_{i}$ is a basis set of one-particle Hilbert space $\mathcal{H}_1$.

The $f$-orbitals can be expanded by $\{\hat{c}^{\dagger}_i\}_{i}$ as
\begin{align}
	\hat{f}^{\dagger}_{\alpha} = \sum_{i} \hat{c}_{i}^{\dagger} u_{i \alpha},
\end{align}
where $\sum_{i} u^{*}_{i \beta}u_{i \alpha} = \delta_{\alpha \beta}$ to be orthonormal.
Therefore, we have
\begin{align}
	P_{\mathcal{W}_f} = \sum_{\alpha} |f_\alpha\rangle \langle f_\alpha| = \sum_{\alpha} \sum_{i,j} u_{i \alpha} u^*_{j \alpha} |i\rangle \langle j|.
\end{align}

We also introduce the orthonormal basis set of the orthogonal complement of $\mathcal{W}_f$, denoted by $\{|g_{\beta}\rangle \coloneqq \hat{g}^{\dagger}_{\beta} | 0 \rangle\}_{\beta}$, whose orbitals are defined by
\begin{align}
	\hat{g}^{\dagger}_{\beta} = \sum_{i} \hat{c}_{i}^{\dagger} v_{i \beta},
\end{align}
where $\sum_{i} v^{*}_{i \beta}v_{i \alpha} = \delta_{\alpha \beta}$ and $\sum_{i} u^{*}_{i \beta}v_{i \alpha} = 0$.
Since $\mathcal{H}_1 = \mathcal{W}_f \oplus \mathcal{W}^{\perp}_{f}$, the creation operator $\hat{c}_{i}^{\dagger}$ can be represented as
\begin{align}
	\hat{c}_{i}^{\dagger} = \sum_{\alpha} u^{*}_{i \alpha} \hat{f}^{\dagger}_{\alpha} + \sum_{\beta} v^{*}_{i \beta} \hat{g}^{\dagger}_{\beta}.
\end{align}
The two-point correlation function can be calculated as
\begin{equation}
	\begin{aligned}
		\langle \Phi | \hat{c}^{\dagger}_{j} \hat{c}_{i} | \Phi \rangle &= \sum_{\alpha, \beta} u^{*}_{j \alpha} u_{i \beta}  \langle \Phi | \hat{f}^{\dagger}_{\alpha} \hat{f}_{\beta} | \Phi \rangle + \sum_{\alpha, \beta} v^{*}_{j \alpha} v_{i \beta} \langle \Phi | \hat{g}^{\dagger}_{\alpha} \hat{g}_{\beta} | \Phi \rangle \\
		&+ \sum_{\alpha, \beta} u^{*}_{j \alpha} v^*_{i \beta} \langle \Phi | \hat{f}^{\dagger}_{\alpha} \hat{g}_{\beta} | \Phi \rangle + \sum_{\alpha, \beta} v^{*}_{j \alpha} u^*_{i \beta} \langle \Phi | \hat{g}^{\dagger}_{\alpha} \hat{f}_{\beta} | \Phi \rangle. \\
		&= \sum_{\alpha} u_{i \alpha} u^{*}_{j \alpha},
	\end{aligned}
\end{equation}
where we use
\begin{align}
	\langle \Phi | \hat{f}^{\dagger}_{\alpha} \hat{f}_{\beta} | \Phi \rangle = \delta_{\alpha \beta},
	\langle \Phi | \hat{g}^{\dagger}_{\alpha} \hat{g}_{\beta} | \Phi \rangle =
	\langle \Phi | \hat{f}^{\dagger}_{\alpha} \hat{g}_{\beta} | \Phi \rangle = 0.
\end{align}
Therefore, we conclude
\begin{align}
	[P_{\mathcal{W}_f}]_{ij} = \langle i| P_{\mathcal{W}_f} |j \rangle =  \sum_{\alpha} u_{i \alpha} u^{*}_{j \alpha} = \langle \Phi | \hat{c}^{\dagger}_{j} \hat{c}_{i} | \Phi \rangle.
\end{align}

\section{Gosset--Huang inequality for fermions}
\label{app:GH}

\subsection{Settings}
Let us consider a frustration-free Hamiltonian $\hat H = \sum_{\bm{R} \in \Lambda} \hat H_{\bm{R}}$  defined on a $d$-dimensional lattice $\Lambda$. 
Without loss of generality, we assume all $\hat H_{\bm{R}}$ are positive-semidefinite and have at least one zero-eigenvalue.
We consider both bosonic and fermionic systems.
In the case of fermionic systems, each $\hat H_{\bm{R}}$ is assumed to be bosonic, which means $[\hat H_{\bm{R}}, (-1)^{\hat F}] = 0$ where $(-1)^{\hat F}$ is the fermion parity.
Fermionic local terms are excluded because physical interaction terms should be local, i.e., sufficiently distant local Hamiltonians commute (not anticommute) with each other.

\subsection{Interaction graph}
\begin{figure}[t]
    \centering
    \includegraphics[width=0.6\columnwidth]{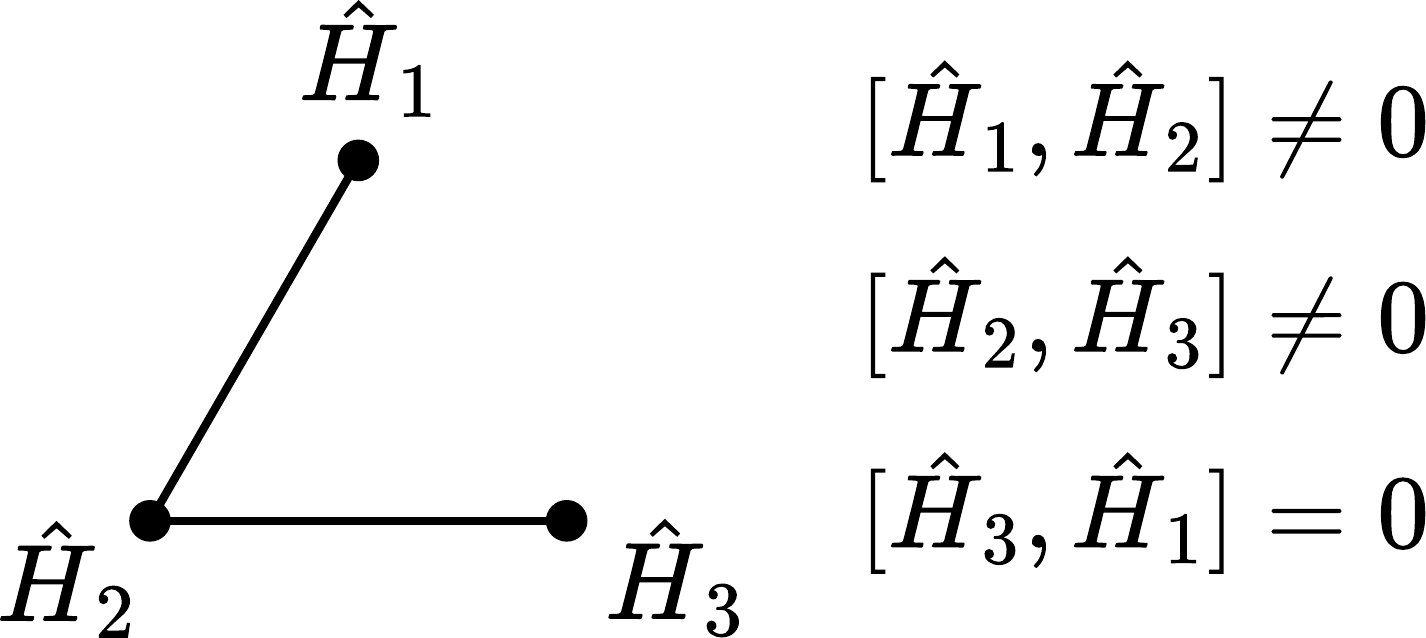}
    \caption{An example of the interaction graph with three vertices. The vertices correspond to local Hamiltonians $\hat H_{\bm{R}}$ and the edges represent the noncommutativity between the local Hamiltonians.}
    \label{fig:Interaction graph}
\end{figure}

For the formulation and the proof of the Gosset--Huang inequality, it is useful to define the interaction graph depicted in Fig.~\ref{fig:Interaction graph}.
The interaction graph is defined for a decomposition $\hat H = \sum_{\bm{R} \in \Lambda} \hat H_{\bm{R}}$ of a Hamiltonian $\hat H$, where $\Lambda$ is the index set consisting of the lattice sites and other degrees of freedom.
The interaction graph for this Hamiltonian has $|\Lambda|$ vertices labeled by $\hat{H}_{\bm{R}}$ ($\bm{R} \in \Lambda$) and there is an edge between vertex $\hat{H}_{\bm{R}}$ and vertex $\hat{H}_{\bm{R}'}$ when $[\hat H_{\bm{R}},\hat H_{\bm{R}'}] \ne 0$.
If two vertices $\hat H_{\bm{R}}$ and $\hat H_{\bm{R}'}$ share an edge, we say that $\hat H_{\bm{R}}$ is adjacent to $\hat H_{\bm{R}'}$.
The degree $g_{\bm{R}}$ of a vertex $\hat{H}_{\bm{R}}$ is given by the number of vertices adjacent to $\hat{H}_{\bm{R}}$.
The maximum degree $g$ of the graph is defined by $g \coloneqq\max_{\bm{R}} g_{\bm{R}}$. Thus, $g$ is defined by
\begin{align}
	g \coloneqq \max_{\bm{R} \in \Lambda} \Big|\{\hat H_{\bm{R}'} \mid [\hat H_{\bm{R}'}, \hat H_{\bm{R}}] \ne 0 \}\Big|.
\end{align}
We only consider the case where the maximum degree $g$ is finite and independent of the system size in the large system limit. 

We define another integer $c$ that characterizes the interaction graph, known as the chromatic number.
The chromatic number $c$ of the interaction graph is the smallest number of colors needed for coloring the vertices so that no two adjacent vertices have the same color.
The chromatic number $c$ is also finite if the maximum degree $g$ is finite since the greedy coloring algorithm guarantees $c \leq g+1$.

The interaction graph offers a natural definition of the distance between two local operators. 
The distance $d(\hat H_{\bm{R}}, \hat H_{\bm{R}'})$ between two vertices $\hat{H}_{\bm{R}}$ and $\hat{H}_{\bm{R}'}$ of the interaction graph is given by the length of the shortest path connecting the two vertices.
In other words, the distance between $\hat{H}_{\bm{R}}$ and $\hat{H}_{\bm{R}'}$ is measured by the shortest length $l$ of a chain of local Hamiltonians
\begin{align}
	\hat H_{\bm{R}_0}, \hat H_{\bm{R}_1}, \cdots, \hat H_{\bm{R}_l},
\end{align}
where $\hat H_{\bm{R}_0} = \hat H_{\bm{R}}$, $\hat H_{\bm{R}_l} = \hat H_{\bm{R}'}$, and $[\hat H_{\bm{R}_i}, \hat H_{\bm{R}_{i+1}}] \ne 0$ for $i = 0,1,\cdots,l-1$.
The distance between general operators $\hat{\mathcal{O}}$ and $\hat{\mathcal{O}}'$ is defined by
\begin{align}
	&D(\hat{\mathcal{O}}, \hat{\mathcal{O}}')\notag\\
	& \coloneqq \min \big\{ d(\hat H_{\bm{R}}, \hat H_{\bm{R}'}) \mid [\hat H_{\bm{R}}, \hat{\mathcal{O}}] \ne 0, [\hat H_{\bm{R}'}, \hat{\mathcal{O}}'] \ne 0\big\}.
\end{align}
Strictly speaking, $D(\hat{\mathcal{O}}, \hat{\mathcal{O}}')$ is not a distance in the mathematical sense because it does not satisfy the triangle inequality.
We also note that the commutator $[\hat{\mathcal{O}}, \hat{\mathcal{O}}']$ does not appear in the definition of the distance, and thus the nonzero commutator $[\hat{\mathcal{O}}, \hat{\mathcal{O}}'] \ne 0$ does not necessarily imply a short distance between these operators.
Therefore, the distance $D(\hat{\mathcal{O}}, \hat{\mathcal{O}'})$ also works for the case that $\hat{\mathcal{O}}$ and $\hat{\mathcal{O}}'$ are fermionic local operators.

\subsection{The Gosset--Huang inequality}
\begin{thm}[Gosset and Huang \cite{gossetCorrelationLengthGap2016}]
 Let $\hat H$ be a frustration-free Hamiltonian with the decomposition $\hat H = \sum_{\bm{R} \in \Lambda} \hat H_{\bm{R}}$ where each $\hat H_{\bm{R}}$ is a positive semidefinite operator 
 Let $\hat G$ be the orthogonal projector onto the ground space and $\epsilon$ be the spectral gap.
 For operators $\hat{\mathcal{O}}$ and $\hat{\mathcal{O}}'$ , the following inequality holds.
 \begin{align}&
 \frac{|\langle\Phi|\hat{\mathcal{O}}(\hat{\mathbbm{1}}-\hat G)\hat{\mathcal{O}}'|\Phi\rangle|}{\|\hat{\mathcal{O}}^\dagger|\Phi\rangle\| \|\hat{\mathcal{O}}'|\Phi\rangle\|} \nonumber\\
 &
 \leq 2e^2 \exp\left\lparen-\frac{D(\hat{\mathcal{O}},\hat{\mathcal{O}}')-1}{c-1}\sqrt{\frac{\epsilon}{g^2\max_{\bm{R}}\|\hat{H}_{\bm{R}}\|}+\epsilon}\right\rparen.\label{GHineq}
 \end{align}
 The definitions of the maximum degree $g$, the chromatic number $c$, and the distance $D(\hat{\mathcal{O}},\hat{\mathcal{O}}')$ between $\hat{\mathcal{O}}$ and $\hat{\mathcal{O}}'$ are given in the previous subsection.
\end{thm}

The proof of the Gosset--Huang inequality is in the original paper~\cite{gossetCorrelationLengthGap2016} and Ref.~\cite{arXiv:2406.06415}.
Although the Gosset--Huang inequality is originally derived for bosonic systems, the extension to fermionic systems with bosonic local Hamiltonians is straightforward.

To obtain a nontrivial inequality, $\hat{\mathcal{O}}$ and $\hat{\mathcal{O}}'$ must be local in the sense that $[\hat{\mathcal{O}},\hat H_{\bm{R}}] = 0$ for sufficiently distant $\hat{\mathcal{O}}$ and $\hat H_{\bm{R}}$ (and similarly for $\hat{\mathcal{O}}'$).
However, the assumption that $\hat{\mathcal{O}}$ and $\hat{\mathcal{O}}'$ commute is not necessary.
Therefore we can use the Gosset--Huang inequality to correlation functions of fermionic operators that anticommute with each other.
	
Now, suppose that there exist operators $\hat{\mathcal{O}}$, $\hat{\mathcal{O}}'$, separated by a distance of order the system size $L$, such that their correlation function shows the power-law decay
\begin{align}
	\frac{|\langle\Phi|\hat{\mathcal{O}}(\hat{\mathbbm{1}}-\hat G)\hat{\mathcal{O}}'|\Phi\rangle|}
	{\|\hat{\mathcal{O}}^\dagger|\Phi\rangle\| \|\hat{\mathcal{O}}|\Phi\rangle\|} \geq CL^{-\Delta} \label{assumption for correlation functions fermion},
\end{align}
where $C$ and $\Delta$ are positive constants.
Then it follows by the Gosset--Huang inequality \eqref{GHineq} that
\begin{align}
	CL^{-\Delta} &\leq \frac{|\langle\Phi|\hat{\mathcal{O}}(\hat{\mathbbm{1}}-\hat G)\hat{\mathcal{O}}'|\Phi\rangle|}{\|\hat{\mathcal{O}}^\dagger|\Phi\rangle\| \|\hat{\mathcal{O}}'|\Phi\rangle\|} 
	\lesssim  \exp\left\lparen -O(L\sqrt\epsilon) \right\rparen.
\end{align}
This implies an upper bound on the finite-size gap as
\begin{align}
	\epsilon \leq C' \frac{(\log L)^2}{L^2},
	\label{eq:upper bound for spectral gap}
\end{align}
where $C'$ is a positive constant, which prohibits the dispersion relation $\omega_{\bm{k}} \sim |\bm{k}|^z$ of low-energy modes with $z < 2$.
We emphasize that our assumption concerns power-law decay as a function of $L$ rather than $D(\hat{\mathcal{O}}_{\bm{x}}, \hat{\mathcal{O}}'_{\bm{y}})$.
Therefore a correlation function like $1/L^{\Delta}$ that does not depend on $D(\hat{\mathcal{O}}, \hat{\mathcal{O}}')$ is sufficient to apply the above argument  and we will see an example in Eq.~\eqref{oneoverLexample}. 

\section{Power-law decaying correlation functions
\label{app:power-law}
}

In this Appendix, we provide two examples of power-law decaying correlation functions satisfying the assumption in Eq.~\eqref{assumption for correlation functions fermion}. We also provide a general argument proving that singular band touching points lead to power-law decaying correlation functions.

\subsection{One-band model}
Let us see a power-law decaying correlation function for the one-band model given by
\begin{align}
	\hat H = \sum_{i=1}^L (\hat c^\dagger_{i+1}\hat c_{i+1} + \hat c^\dagger_{i}\hat c_{i} - \hat c^\dagger_{i+1}\hat c_{i} -\hat c^\dagger_{i}\hat c_{i+1}).
\end{align}
We can check the frustration-freeness of this model by the following decomposition.
\begin{align}
	\hat H_i = \sum_{i=1}^L \hat H_i,\quad\hat H_i = \hat{\psi}^\dagger_i\hat{\psi}_i, \quad
	\hat{\psi}_i = \hat{c}_{i+1} - \hat{c}_i.
\end{align}
This model is gapless with the quadratic dispersion relation $\omega_k = 2(1-\cos k)$.
We see this property from the power-law decaying correlation.
The ground states are given by
\begin{align}
	|0\rangle, \quad
	|\Phi_1\rangle \coloneqq \frac{1}{\sqrt{L}}\sum_i \hat{c}_i^\dagger |0\rangle,
\end{align}
where $|0\rangle$ is the Fock vacuum state.
We consider the correlation function between $\hat{c}_i$ and $\hat{c}^\dagger_j$ for $i \ne j$ on the ground state $|0\rangle$.
The calculation is given by
\begin{align}
	\frac{\langle 0 | \hat{c}_i (\hat{\mathbbm{1}}-\hat{G} )\hat{c}_j^\dagger |0\rangle}{\|\hat c^\dagger_i |0\rangle\|  \|\hat c^\dagger_j |0\rangle\|}
	&
	= \langle 0 | \hat{c}_i (\hat{\mathbbm{1}}-\hat{G} )\hat{c}_j^\dagger |0\rangle \nonumber\\
	&
	= -\langle0|\hat c_i|\Phi_1\rangle\langle\Phi_1|\hat c^\dagger_j|0\rangle \nonumber\\
	&
	= -\frac1{L}\sum_{k,l}\langle0|\hat c_i\hat c^\dagger_k|0\rangle\langle0|\hat c_l\hat c^\dagger_j|0\rangle \nonumber\\
	&
	= -\frac1{L}.\label{oneoverLexample}
\end{align}
Here we used the fact that
\begin{align}
	\langle0|\hat{c}_i\hat{\mathbbm{1}}\hat{c}^\dagger_j|0\rangle
	= \langle0|\hat{c}_i |0\rangle \langle0|\hat{c}^\dagger_j|0\rangle = 0.
\end{align}
Let $j > i$ without loss of generality. The distance between $\hat{c}_i$ and $\hat{c}^\dagger_j$ is given by
\begin{align}
	D(\hat{c}_i, \hat{c}^\dagger_j) = d(\hat H_{i}, \hat H_{j-1}) = j-i-1.
\end{align}
By taking $|i-j| \sim L$, the assumption in Eq.~\eqref{assumption for correlation functions fermion} is satisfied and we obtain the upper bound on the finite-size gap in Eq.~\eqref{eq:upper bound for spectral gap}.

\subsection{Checkerboard lattice}
Next, we consider the model on the checkerboard lattice discussed in the main text.
For simplicity, we move the origin of momentum from $(0, 0)$ to $(\pi, \pi)$ so that the zero-energy mode is located at $\bm{k} = (0, 0)$.
In other words, we redefine the fermion operators as
\begin{align}
	\hat a_{\bm{R}} \to (-1)^{R_x+R_y}\hat a_{\bm{R}},\quad
	\hat b_{\bm{R}} \to (-1)^{R_x+R_y}\hat b_{\bm{R}}.
\end{align}
Then, the unnormalized Bloch state for the occupied band in Eq.~\eqref{CBBloch} is replaced by
\begin{align}
	\bm{w}_{\bm{k}} = \left\lparen\begin{matrix} 
		1-e^{-ik_x} \\
		-1+e^{-ik_y}
	\end{matrix}\right\rparen.
\end{align}
The zero-energy modes are generated by $\hat a_{\bm{k} = \bm{0}}^\dagger$ and $\hat b_{\bm{k} = \bm{0}}^\dagger$.
The ground states are given by the following four states.
\begin{align}&
	|\Phi\rangle \coloneqq \prod_{\bm{k} \ne \bm{0}} (w_{\bm{k}1}\hat a^\dagger_{\bm{k}} + w_{\bm{k}2}\hat b^\dagger_{\bm{k}})|0\rangle, \\
	&
	|\Phi_a\rangle \coloneqq \hat a^\dagger_{\bm{k}=\bm{0}}|\Phi\rangle, \\
	&
	|\Phi_b\rangle \coloneqq \hat b^\dagger_{\bm{k}=\bm{0}}|\Phi\rangle, \\
	&
	|\Phi_{ab}\rangle \coloneqq \hat a^\dagger_{\bm{k}=\bm{0}}\hat b^\dagger_{\bm{k}=\bm{0}}|\Phi\rangle.
\end{align}
Let us compute the following correlation function.
\begin{align}
	\frac{\langle\Phi|\hat a_{\bm{0}}(\hat{\mathbbm{1}}-\hat G)a^\dagger_{\bm{R}}|\Phi\rangle}{\| a^\dagger_{\bm{0}}|\Phi\rangle\| \| a^\dagger_{\bm{R}}|\Phi\rangle\|}.
\end{align}
Correlation functions involving ground states other than $|\Phi\rangle$, as well as those involving $\hat b_{\bm{R}}$ instead of $\hat a_{\bm{R}}$, can be computed  similarly as follows.
The distance between $\hat{a}_{\bm{0}}$ and $\hat{a}^\dagger_{\bm{R}}$ is given by $|R_x| + |R_y| - 1$.
The contribution from the projector $\hat G$ onto the ground space is written as
\begin{align}&
	\langle\Phi|\hat a_{\bm{0}}\hat G\hat a^\dagger_{\bm{R}}|\Phi\rangle \nonumber\\
	&
	= \langle\Phi|\hat a_{\bm{0}}|\Phi\rangle\langle\Phi|\hat a^\dagger_{\bm{R}}|\Phi\rangle 
	+ \langle\Phi|\hat a_{\bm{0}}|\Phi_a\rangle\langle\Phi_a|\hat a^\dagger_{\bm{R}}|\Phi\rangle \nonumber\\
	&
	+ \langle\Phi|\hat a_{\bm{0}}|\Phi_b\rangle\langle\Phi_b|\hat a^\dagger_{\bm{R}}|\Phi\rangle 
	+ \langle\Phi|\hat a_{\bm{0}}|\Phi_{ab}\rangle\langle\Phi_{ab}|\hat a^\dagger_{\bm{R}}|\Phi\rangle.
\end{align}
The first term and the last term vanish because of the conservation of the fermion number. The third term also vanishes because 
\begin{align}
	\langle\Phi_b|\hat a^\dagger_{\bm{R}}|\Phi\rangle = \langle\Phi|\hat b_{\bm{k}=\bm{0}}\hat a^\dagger_{\bm{R}}|\Phi\rangle = -\langle\Phi|\hat a^\dagger_{\bm{R}}\hat b_{\bm{k}=\bm{0}}|\Phi\rangle = 0
\end{align}
The second term is computed as $1/V$ using the following calculation.
\begin{align}
	\langle\Phi_a|\hat a^\dagger_{\bm{R}}|\Phi\rangle
	&
	= \frac1{V} \sum_{\bm{R}} \langle\Phi_a|\hat a^\dagger_{\bm{R}}|\Phi\rangle \nonumber\\
	&
	= \frac1{\sqrt{V}}\langle\Phi_a|\hat a^\dagger_{\bm{k}=\bm{0}}|\Phi\rangle 
	= \frac1{\sqrt{V}}.
\end{align}
The contribution from the identity $\hat{\mathbbm{1}}$ is described as
\begin{align}
	\langle\Phi|\hat a_{\bm{0}}\hat{\mathbbm{1}}\hat a^\dagger_{\bm{R}}|\Phi\rangle
	&
	= \frac1{V} \sum_{\bm{k}} e^{i\bm{k}\cdot\bm{R}} \|\hat a^\dagger_{\bm{k}}|\Phi\rangle \|^2 \nonumber\\
	&
	= \frac1{V} \sum_{\bm{k}} e^{i\bm{k}\cdot\bm{R}} (P_\mathrm{unoc}(\bm{k}))_{11},
	\label{identity part of correlation on checker board}
\end{align}
where $P_{\mathrm{unoc}}(\bm{k})$ is a projector onto the vector space spanned by the Bloch states of the unoccupied bands at the momentum $\bm{k}$.
This is computed for $\bm{k} \ne \bm{0}$ as
\begin{align}&
	P_\mathrm{unoc}(\bm{k}) = \mathbbm{1} - \frac{\bm{w}_{\bm{k}}\bm{w}^\dagger_{\bm{k}}}{\|\bm{w}_{\bm{k}}\|^2} \nonumber\\
	&
	= \frac1{4-2\cos k_x-2\cos k_y} \nonumber\\
	&
	\times \left\lparen\begin{matrix} 
		2-2\cos k_y & (1-e^{-ik_x})(1-e^{ik_y}) \\
		(1-e^{ik_x})(1-e^{-ik_y}) & 2-2\cos k_x
	\end{matrix}\right\rparen.
\end{align}
For $\bm{k} = \bm{0}$, we have $P_\mathrm{unoc}(\bm{k}=\bm{0}) = \mathbbm{1}$ from the definition of the ground state $|\Phi\rangle$.
Therefore, the integrand of the Fourier transformation in Eq.~\eqref{identity part of correlation on checker board} is given by
\begin{align}
	\|\hat a^\dagger_{\bm{k}}|\Phi\rangle\|^2
	= \frac{1-\cos k_y}{2 - \cos k_x - \cos k_y}
	+ \frac{\delta(\bm{k})}{V}.
\end{align}
Here, the delta function is defined as
\begin{align}
	\delta(\bm{k}) = \begin{cases}
		V & (\bm{k} = \bm{0}), \\
		0 & (\bm{k} \ne \bm{0}).
	\end{cases}
\end{align}
The contribution from the delta function is canceled by the contribution from the projector $\hat G$.
Using continuum approximation, the unnormalized correlation function is computed as
\begin{align}
	\langle\Phi|\hat a_{\bm{0}}(\hat{\mathbbm{1}}-\hat G)\hat a^\dagger_{\bm{R}}|\Phi\rangle
	&
	= \frac1{V}\sum_{\bm{k}} {{e}}^{{{i}}\bm{k}\cdot\bm{R}} \frac{1-\cos k_y}{2 - \cos k_x - \cos k_y} \nonumber\\
	&
	\approx \int\frac{d^2\bm{k}}{(2\pi)^2} {{e}}^{{{i}}\bm{k}\cdot\bm{R}} \frac{k_y^{\,2}}{k_x^{\,2}+k_y^{\,2}} \nonumber\\
	&
	= \partial_y^{\,2}\left\lparen \frac1{2\pi}\ln|\bm{R}| \right\rparen \nonumber\\
	&
	= \frac1{2\pi}\frac{R_y^{\,2} - R_x^{\,2}}{|\bm{R}|^4}.
\end{align}
Here, we used the fact that the Green function of the Laplacian operator in two dimensions is given by $(1/2\pi)\ln|\bm{R}|$. Since the normalization factor satisfies $\|\hat a^\dagger_{\bm{0}}|\Phi\rangle\|\|\hat a^\dagger_{\bm{R}}|\Phi\rangle\| = \langle\Phi|a_{\bm{R}}a^\dagger_{\bm{R}}|\Phi\rangle \approx 1/2$, we obtain
\begin{align}
	\frac{|\langle\Phi|\hat a_{\bm{0}}(\hat{\mathbbm{1}}-\hat G)\hat a^\dagger_{\bm{R}}|\Phi\rangle|}{\|\hat a^\dagger_{\bm{0}}|\Phi\rangle\|\|\hat a^\dagger_{\bm{R}}|\Phi\rangle\|} \sim \frac1{L^2},
\end{align}
by taking $D(\hat a_{\bm{0}}, \hat a^\dagger_{\bm{R}}) \sim L$.
Therefore, the upper bound on the finite-size gap in Eq.~\eqref{eq:upper bound for spectral gap} also holds for this case.
The same algebraic dependence of the correlation function can be obtained for the model on the Kagome lattice discussed in Sec.~\ref{sec:examples}.

\subsection{General argument for singular band touching points}

In the preceding calculations, we observed that the long-range behavior of the correlation function differs significantly depending on the model, with some exhibiting a power-law decay with distance.
This long-range behavior is distinct from finite-size effects, which scale with the overall system size.
This difference can be systematically understood by characterizing the nature of the band touching points. We define a band touching point as singular if the quantum distance between nearby Bloch states remains finite as the touching point is approached~\cite{rhimSingularFlatBands2021}. In this section, we provide the general discussion that it is precisely this singularity that gives rise to the power-law correlations observed previously.

The quantum distance between the Bloch states $w_{\bm{k}}$ is defined by
\begin{align}
    d(\bm{k}, \bm{k}') \coloneqq 1 - \frac{w_{\bm{k}}^\dagger w_{\bm{k}'}}{\|w_{\bm{k}}\|\|w_{\bm{k}'}\|}.
\end{align}
Typically, $d(\bm{k}, \bm{k}') \to 0$ as $|\bm{k}-\bm{k}'| \to 0$, but it can remain finite in the vicinity of a singular band touching point.

The correlation function under consideration is
\begin{align}
    \langle \Phi | \hat{a}_{\bm{0}}(\hat{\mathbbm{1}}-\hat{G})\hat{a}^\dagger_{\bm{R}} | \Phi \rangle.
\end{align}
Here, the normalization constants $\|\hat{a}_{\bm{0}}| \Phi \rangle\|$ and $\|\hat{a}^\dagger_{\bm{R}}|\Phi\rangle\|$ are omitted as they do not alter the scaling of the correlation function.
The contribution from the identity operator is given by
\begin{align}
    \langle\Phi|\hat{a}_{i,\bm{0}}\hat{\mathbbm{1}}\hat{a}^\dagger_{j,\bm{R}}|\Phi\rangle
    &= \frac{1}{V} \sum_{\bm{k}} e^{i\bm{k}\cdot\bm{R}}  \langle\Phi|\hat{a}_{i,\bm{k}}\hat{a}^\dagger_{i,\bm{k}}|\Phi\rangle \nonumber\\
    &= \frac{1}{V} \sum_{\bm{k}} e^{i\bm{k}\cdot\bm{R}} (P_\mathrm{unoc}(\bm{k}))_{ii},
    \label{identity contribution}
\end{align}
where $P_\mathrm{unoc}(\bm{k})$ is the projection operator onto the unoccupied bands at momentum $\bm{k}$.
It should be noted that for a frustration-free system, the positive and negative energy bands are completely separated, except at band touching points.

For simplicity, we consider systems with one singular band touching point at $\bm{k} = \bm{0}$. In such systems, the projector $P_\mathrm{unoc}(\bm{k})$ exhibits singular behavior near $\bm{k} = \bm{0}$. However, if the singular band touching consists entirely of positive-energy bands or entirely of negative-energy bands, the singularity in $P_\mathrm{unoc}(\bm{k})$ is canceled. Furthermore, there is a degree of freedom in $| \Phi \rangle$ regarding the excitation of the $\bm{k}=\bm{0}$ mode. This introduces a correction to $P_\mathrm{unoc}(\bm{k})$ that is proportional to $\delta(\bm{k})/V$.

The contribution from the projection onto the ground state, $\langle \Phi |\hat{a}_{\bm{0}}\hat{G}\hat{a}^\dagger_{\bm{R}}| \Phi \rangle$, provides an $\bm{R}$-independent term of the order of $\delta(\bm{k})/V$. If this term does not perfectly cancel the delta function arising from the identity operator's contribution, it results in a finite-size effect that is independent of the distance $\bm{R}$. It is noteworthy that while such a finite-size effect in the correlation function is a general property not restricted to singular band touching, it is sufficient to guarantee the gaplessness of the system via the Gosset-Huang inequality.

Since our interest lies in the spatial dependence of the correlation function, we return to the evaluation of Eq.~\eqref{identity contribution}.
Let $w_{\alpha,i}(\bm{k})$ be a Bloch state where $\alpha$ labels the band index and $i$ denotes the internal degrees of freedom of the fermions.
Near the touching point $\bm{k} = \bm{0}$, the projection onto the unoccupied bands is expressed using the Bloch states as
\begin{align}
    (P_\mathrm{unoc}(\bm{k}))_{ii} = \sum_{\alpha \in \mathrm{unoc}} \frac{w_{\alpha,i}^*(\bm{k}) w_{\alpha,i}(\bm{k})}{\|w_\alpha(\bm{k})\|^2} \approx g(\Omega_{\bm{k}}) + O(|\bm{k}|),
\end{align}
where the function $g(\Omega_{\bm{k}})$ depends only on the solid angle $\Omega_{\bm{k}}$ of the momentum vector $\bm{k}$, and not on its magnitude $|\bm{k}|$. If the band touching point is not singular, meaning it does not provide a finite quantum distance around the touching point, $g(\Omega_{\bm{k}})$ becomes a direction-independent constant.

The long-range behavior of the correlation function is determined by the Fourier transform of $g(\Omega_{\bm{k}})$. For sufficiently large $|\bm{R}|$, the dominant contribution to the integral comes from a small sphere of radius $|\bm{k}| \sim O(|\bm{R}|^{-1})$. Therefore, the integration domain can be extended from the Brillouin zone to all of momentum space.
Using polar coordinates $\bm{k} = (k, \Omega_{\bm{k}})$ and $\bm{R} = (R, \Omega_{\bm{r}})$, we have $d^dk = k^{d-1}dkd\Omega_{\bm{k}}$ and $\bm{k}\cdot\bm{R} = kR\cos \theta$. The integral can thus be written as:
\begin{align}
    \langle\Phi|\hat{a}_{i,\bm{0}}\hat{a}^\dagger_{j,\bm{R}}|\Phi\rangle
    &
    \approx \int d\Omega g(\Omega) \int_0^\infty k^{d-1}dk\; e^{i\bm{k}\cdot\bm{R}} \nonumber\\
    &
    = R^{-d} \int d\Omega g(\Omega) \int_0^\infty u^{d-1}du\; e^{iu\cos\theta}.
\end{align}
Thus, the correlation function is proportional to $R^{-d}$. However, if $g(\Omega)$ is direction-independent, the integral vanishes for $\bm{R}\neq\bm{0}$. This implies that the power-law decay of the ground-state correlation function is a characteristic feature of singular band touching points.

\bibliography{refs}
\end{document}